\documentclass[12pt,twocolumn]{emulateapj}

\tightenlines

\usepackage{ulem} 
\usepackage{color} 
\usepackage{graphicx}
\usepackage{epsfig,amsmath}
\usepackage{times}

\makeatletter

\newcommand{\Rmnum}[1]{\expandafter\@slowromancap\romannumeral #1@}
\makeatother

\newcount\listnorom
\newcommand\listromanDE{\global\advance \listnorom by 1
{\lowercase\expandafter{(\romannumeral\listnorom)}\ }}

\def\lsim{\raise0.3ex
  \hbox{$<$\kern-0.75em\raise-1.1ex\hbox{$\sim$}}\,}
\def\gsim{\raise0.3ex
  \hbox{$>$\kern-0.75em\raise-1.1ex\hbox{$\sim$}}\,}

\newcommand{\be}{\begin{eqnarray}}
\newcommand{\ee}{\end{eqnarray}}

\newcommand{\mrm}{\mathrm}

\newcount\Inum
\newcount\IInum
\newcount\IIInum
\newcount\IVnum
\def\I{\global\multiply\IInum by 0 \global\multiply\IIInum by 0
            \global\multiply\IVnum by 0 \global\advance \Inum by 1
            {\the\Inum. }}
\def\II{\global\multiply\IIInum by 0\global\multiply\IVnum by 0
       \global\advance \IInum by 1 {\the\Inum.\the\IInum. }}
\def\III{\global\multiply\IVnum by 0\global\advance \IIInum by 1
            {\the\Inum.\the\IInum.\the\IIInum. }}
\def\IV{\global\advance \IVnum by 1
            {\the\IVnum. }}
\begin{document}

\title{Shock acceleration of electrons and synchrotron emission from the dynamical ejecta of neutron star mergers}

\author{
Shiu-Hang Lee\altaffilmark{1},
Keiichi Maeda\altaffilmark{1},
Norita Kawanaka\altaffilmark{1,2}
}

\altaffiltext{1}{Department of Astronomy, Kyoto University, Kitashirakawa, Oiwake-cho, Sakyo-ku, Kyoto 606-8502, Japan; herman@kusastro.kyoto-u.ac.jp}
\altaffiltext{2}{Hakubi Center for Advanced Research, Kyoto University, Yoshida-honmachi, Sakyo-ku, Kyoto 606-8501, Japan}

\begin{abstract}
Neutron star mergers (NSMs) eject energetic sub-relativistic dynamical ejecta into the circumbinary media. As analogous to supernovae and supernova remnants, the NSM dynamical ejecta are expected to produce non-thermal emission by electrons accelerated at a shock wave. In this paper, we present expected radio and X-ray signals by this mechanism, taking into account non-linear diffusive shock acceleration (DSA) and magnetic field amplification. We suggest that the NSM has a unique nature as a DSA site, where the seed relativistic electrons are abundantly provided by the decays of r-process elements. The signal is predicted to peak at a few 100 - 1,000 days after the merger, determined by the balance between the decrease of the number of seed electrons and the increase of the dissipated kinetic energy due to the shock expansion. While the resulting flux can ideally reach to the maximum flux expected from near-equipartition, the available kinetic energy dissipation rate of the NSM ejecta limits the detectability of such a signal. It is likely that the radio and X-ray emission are overwhelmed by other mechanisms (e.g., an off-axis jet) for an observer placed to a jet direction (i.e., for GW170817). On the other hand, for an off-axis observer, to be discovered once a number of NSMs are identified, the dynamical ejecta component is predicted to dominate the non-thermal emission. While the detection of this signal is challenging even with near-future facilities, this potentially provides a robust probe of the creation of r-process elements in NSMs. 
\end{abstract}


\section{Introduction}

The gravitational waves (GWs) from a binary neutron star merger (NSM) was detected for the first time by \textit{Advanced LIGO} and \textit{VIRGO} \citep{2017PhRvL.119p1101A}.  This event, GW170817, was followed by a short gamma-ray burst GRB 170817A \citep{2017ApJ...848L..14G, 2017ApJ...848L..15S, 2017ApJ...848L..13A} and, thanks to the follow-up observations in NIR, optical, and UV, a luminous and rapidly-evolving electromagnetic counterpart was discovered \citep{Coulter1556, 2017ApJ...848L..12A, 2017Natur.551...64A, 2017ApJ...848L..17C, 2017Natur.551...75S, 2017PASJ...69..101U, 2017PASJ...69..102T, 2017ApJ...848L..24V, Drout1570, Evans1565, 2017ApJ...848L..27T, 2017ApJ...848L..16S, 2017ApJ...848L..18N, 2017Natur.551...67P, 2017ApJ...851L..21V, 2017Natur.551...80K, 2017PASA...34...69A, 2017ApJ...848L..19C, Kilpatrick1583, 2017ApJ...848L..32M, 2017Sci...358.1574S}.  Its observational properties are remarkably consistent with what are expected in the theoretical models of a kilonova/macronova emission powered by radioactive elements in the mildly relativistic ejecta \citep{1998ApJ...507L..59L, 2010MNRAS.406.2650M, 2013ApJ...774...25K, 2013ApJ...775...18B, 2013ApJ...775..113T}.  Later, an afterglow was detected in X-ray and radio bands at $\sim 10$ days after the GW trigger \citep{2017ApJ...848L..20M, 2017Natur.551...71T, 2017ApJ...848L..25H, 2017ApJ...848L..21A, Hallinan1579}.  These components are interpreted as the synchrotron emission from non-thermal electrons, and various models that can explain their early observational behavior have been proposed: a jet seen off-axis \citep{2017ApJ...850L..24G, 2017arXiv171005905I, 2018ApJ...852L...1Z, 2017arXiv171203237L}, a cocoon energized by a jet choked in the ejecta \citep{Kasliwal1559, 2017arXiv171005897B, 2017arXiv171005896G, 2017ApJ...848L..34M}, or an isotropic fireball expanding ahead of the kilonova/macronova ejecta \citep{2017arXiv171103112S}.  Recently this non-thermal component has been found to be still brightening for $> 100$ days, as detected by the long term radio and X-ray observations \citep{2018Natur.554..207M, 2018arXiv180103531M, 2018ApJ...853L...4R}, which may disfavor the interpretation of a homogeneous jet observed off-axis (however, see \citealp{2018arXiv180109712N}.)

Beside these components, we expect the emission from non-thermal electrons accelerated at the shock formed by the kilonova/macronova ejecta interacting with the circumbinary medium (CBM) in radio and/or X-ray bands \citep{2011Natur.478...82N, 2013MNRAS.430.2121P, 2013MNRAS.430.2585R,2014PhRvD..89f3006T, 2015MNRAS.450.1430H, 2016ApJ...831..190H}.  This component is expected to arise at weeks to years after the merger, which is much later than the emission from the relativistic jet.  As for GW170817, the non-thermal radio and X-ray emission from the ejecta-CBM interaction have been calculated using the parameters inferred from the kilonova/macronova emission \citep{2017ApJ...848L..21A, 2018ApJ...852..105A,2018arXiv180300599H}.  However, these calculations are based on the phenomenological parameterization, which for example assumes the constant energy fraction of non-thermal electrons $\varepsilon_e$ and of the magnetic field $\epsilon_B$ to the total dissipated energy at the shock.  In order to predict the non-thermal emission and its energy spectrum from the kilonova/macronova ejecta self-consistently, one should solve the diffusive shock acceleration (DSA) of particles in the ejecta taking into account the modification of the shock structure and magnetic field amplification due to the efficient acceleration of particles.

In this work, we calculate the particle acceleration at the forward shock (FS) in the kilonova/macronova ejecta interacting with the CBM by using the numerical code that couples the DSA of particles and hydrodynamical evolution of the ejecta and collisionless shocks, taking into account the non-linear effect such as the magnetic field amplification and shock modification. This enables us to predict the radio and X-ray emission from non-thermal electrons accelerated at the shock in a more realistic way than previously performed.  Moreover, we consider one physical process that can affect the resulting non-thermal emission from the merger ejecta substantially, which was however not noticed previously. It is the electron injection into the DSA from the decay of radioactive \textit{r}-process nuclei synthesized in the kilonova/macronova ejecta. We show that this process can indeed dominate the seed electrons and the resulting non-thermal emission, and can thus provide new diagnostics of the \textit{r}-process nucleosynthesis in NSMs. We hereafter refer to the kilonova/macronova ejecta simply as the `dynamical ejecta'.

\section{Model}

\begin{figure}
\centering
\includegraphics[width=7cm]{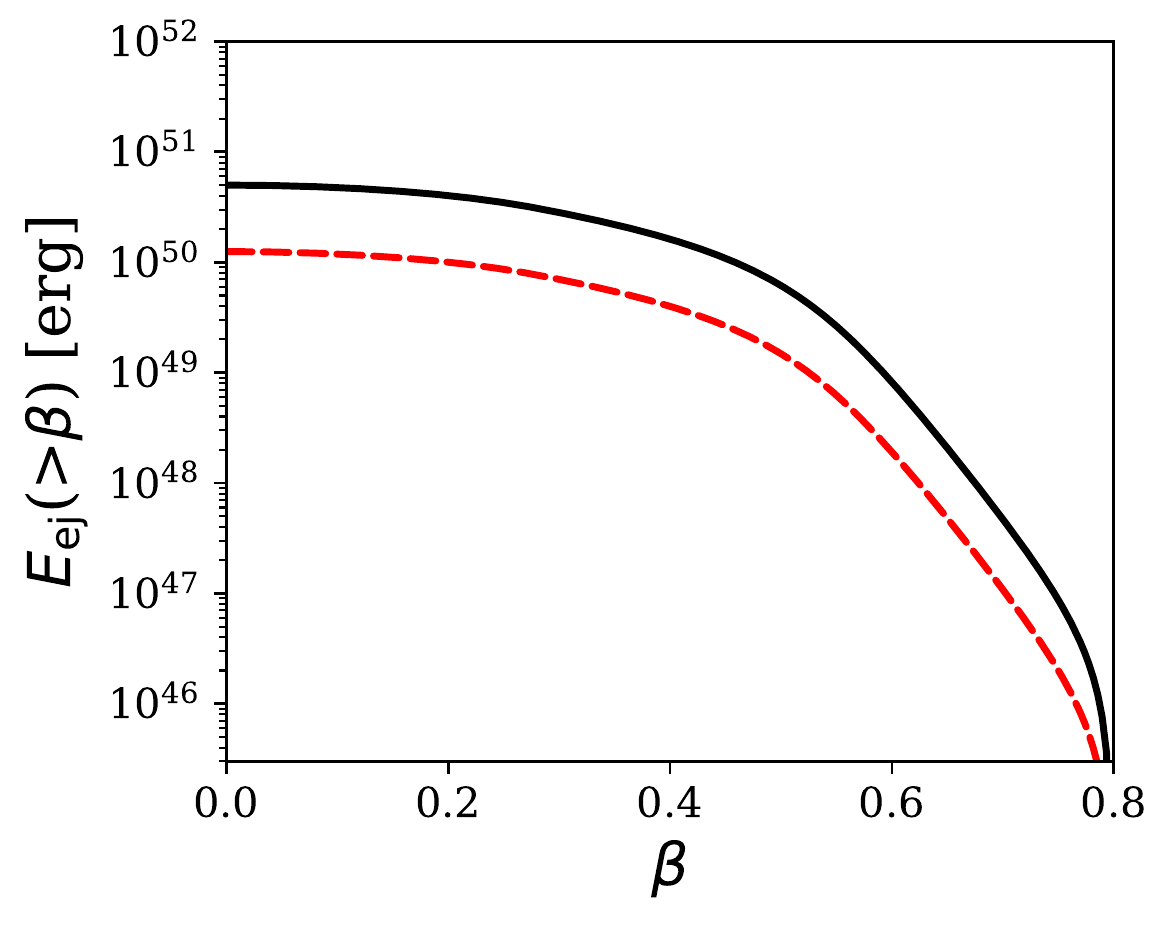}  
\caption{
Initial profiles of cumulative kinetic energy in the NSM ejecta of our models. The black solid line shows the profile for our fiducial models with $E_\mrm{ej} = 5 \times 10^{50}$~erg and $M_\mathrm{ej} = 0.04\ M_\odot$,  and the red dashed line for models with a lowered energetics $E_\mrm{ej} = 1.25 \times 10^{50}$~erg and $M_\mathrm{ej} = 0.01\ M_\odot$ (see text and Table~\ref{table1} for details).
}
\label{fig1}
\end{figure}

In our model, we assume a spherical expansion of the dynamical ejecta into a uniform CBM of constant density $n_\mrm{CBM}$ and magnetic field $B_0$. We adopt an initial density profile for the ejecta proposed by \citet{2016ApJ...831..190H} as follows:
\begin{equation}
\rho_\mrm{ej}(\beta) = \frac{\rho_0 (\beta/\beta_0)^{-\alpha}}{1 + \mrm{exp}[(\beta-2\beta_0)/\sigma]},
\end{equation}
where $\beta \equiv v/c$ is the velocity of the ejecta material, $\alpha = -3$ for $\beta < \beta_0$ and $\alpha = -4.5$ for $\beta \ge \beta_0$ respectively, $\sigma = 0.035$, and $\rho_0$ and $\beta_0$ are parameters set to match the ejecta mass $M_\mrm{ej}$ and kinetic energy $E_\mrm{ej}$ as listed in Table~\ref{table1}. Here $\beta_0 = 0.28$ is adopted. The resulted ejecta kinetic energy profiles of our models are shown in Figure~\ref{fig1}. 

%

\begin{figure}
\centering
\includegraphics[width=6cm]{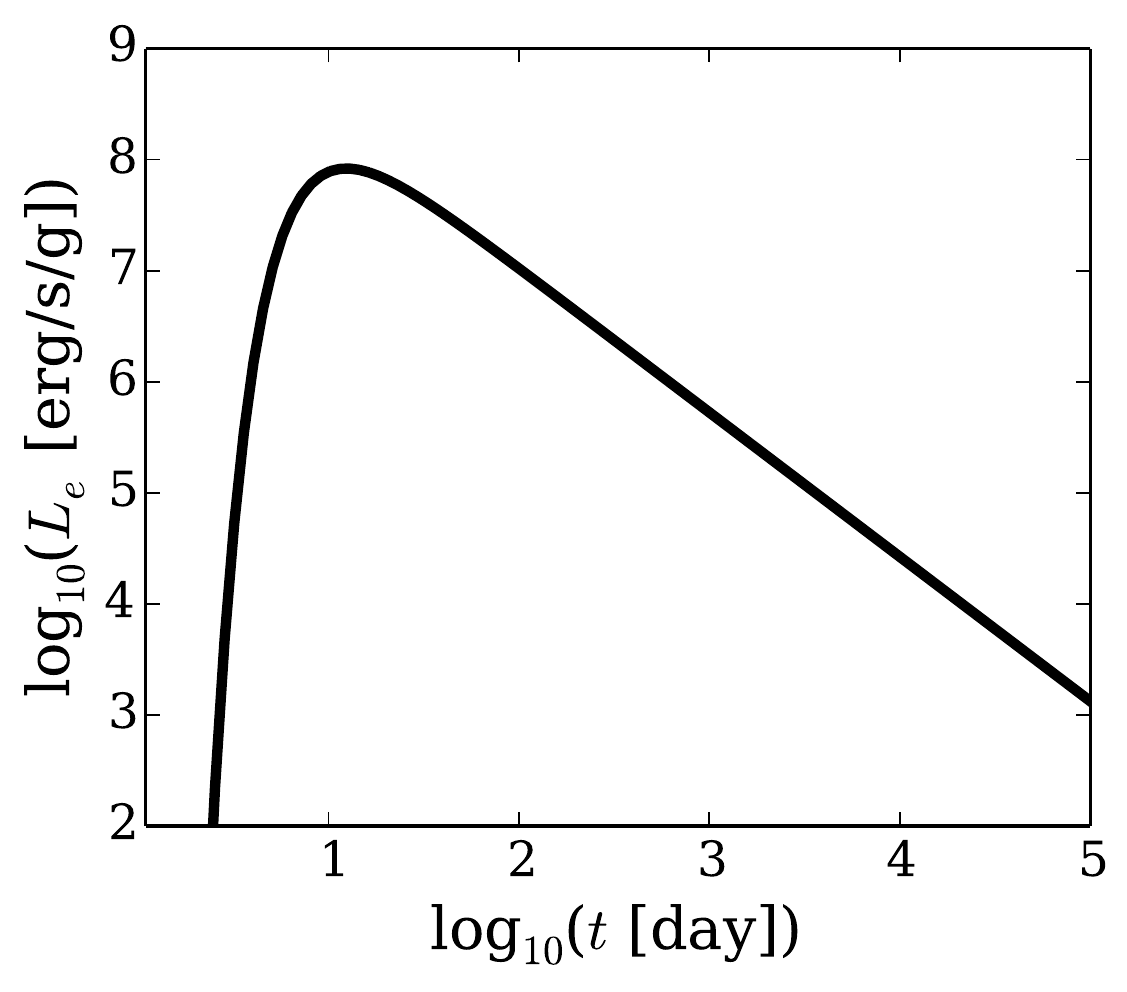}  
\caption{
Time evolution model for the bolometric luminosity per ejecta mass of $\beta$-decay electrons from r-process elements in the dynamical ejecta.
}
\label{fig2}
\end{figure}

Particle acceleration and hydrodynamical evolution of the expanding ejecta are treated by the \textit{CR-Hydro-NEI} Lagrangian hydrodynamic code \citep[e.g.,][]{EPSBG2007,PES2009,PSRE2010,EPSR2010,ESPB2012,LEN2012}. The code couples a non-linear diffusive shock acceleration (NLDSA) of particles with the hydrodynamical evolution of the expanding ejecta and collisionless shockwaves which allows for a self-consistent calculation of multi-wavelength emission properties in spacetime. Magnetic field amplification (MFA) due to streaming instability induced by the accelerating protons in the shock precursor \citep[e.g.,][]{CBAV2009} is taken into account. In this work, we do not consider particle acceleration at the reverse shock. 

The processes as mentioned above are essentially the same with those at work in producing non-thermal emission from supernovae (SNe) and SN remnants (SNRs), for which our code has been tested \citep[e.g.,][]{LSENP2013,Slane2014}. In SNe and SNRs, the injection source for radio and X-ray emitting electrons is thermal particles injected into the DSA process through a thermal leakage mechanism \citep[e.g.,][]{BGV2005}. A typical number fraction $\eta_\mrm{inj,th} \sim 10^{-5}$ to $10^{-4}$ of the shock heated protons with supra-thermal energies are injected and accelerated at the FS. We parameterize the injection of electrons by scaling the accelerated electron spectrum with the proton spectrum through a electron-to-proton number density ratio $K_{ep}$ at relativistic energies. 
While there exist a number of multi-wavelength emission models which support a $K_{ep} \gtrsim 0.001$ for SNRs evolving in a relatively low-density ambient medium, such as Tycho \citep[e.g.,][]{2012A&A...538A..81M, Slane2014}, it is still an open question if the electron injection process at shocks created by NSM ejecta is identical to the case of SNRs. We choose to set $K_{ep} = 0.01$ as our fiducial value, and reserve more detailed discussions for the near future when more abundant observational constraints become available from detections of other NSM events.

Considering the properties of the NSM ejecta, we identify another source of the electron injection, which potentially overwhelms the contribution from the supra-thermal particles; non-thermal electrons emitted by the decay of the neutron-rich radioactive isotopes in the dynamical ejecta. These non-thermal electrons (typical energy of $\sim 100$ keV to $\sim 1$ MeV) are initially thermalized efficiently within the ejecta in the first $\sim 10$ days and contribute to the NIR-optical emission \citep{2013ApJ...775...18B,2016ApJ...829..110B}. Later on, the ejecta become transparent to these electrons, and thus they can travel outwards and be picked up by the FS and further accelerated to higher energies. Here we consider the possibility that a fraction of these $\beta$-decay electrons are injected into DSA and be accelerated in the same manner as the shock heated particles. 

Figure~\ref{fig2} shows our input model for the luminosity evolution of the $\beta$-decay electrons emitted by the r-process elements in the dynamical ejecta per unit ejecta mass, which can be described by the following expression \citep[see Equation 2 of][]{2016MNRAS.459...35H}:
\begin{equation}
L_e(t) = 4.2 \times 10^9\ \left(\frac{t}{{\rm day}}\right)^{-1.3}\ \mrm{exp}\left[-\left(\frac{t}{t_\mrm{esc}}\right)^{-2}\right]\ {\rm erg} \ {\rm s}^{-1} {\rm g}^{-1} \ ,
\end{equation}
where $t$ is time, and $t_\mrm{esc} \sim 10$~day is the typical escape timescale of $\beta$-decay electrons from the ejecta. 
The number density of these electrons at the shock can then be estimated as:
\begin{equation}
n_{e,\mrm{decay}}(t) = \frac{L_e(t')M_\mrm{ej}}{4\pi R_\mrm{sk}^2 \widetilde{v_e} \widetilde{E_e}}, 
\end{equation}
where $R_\mrm{sk}(t)$ is the shock radius, and $\widetilde{v_e}$, $\widetilde{E_e}$ are the typical velocity and energy of the decay electrons respectively. Electrons that escaped from the ejecta at time $t'$ arrive at the FS at time $t$ in the observer frame, so that
\begin{equation}
 t = t' + \left(\frac{R_\mrm{sk}-R_\mrm{esc}}{\widetilde{v_e} - v_\mrm{sk}}\right) \left(1 - \frac{\widetilde{v_e}v_\mrm{sk}}{c^2}\right),
\end{equation} 
where $R_\mrm{esc}$ is the typical radius of the escape surface for the electrons. For simplicity we assume $R_\mrm{esc} \sim R_\mrm{CD}/2$ where $R_\mrm{CD}$ is the contact discontinuity radius.

These electrons are injected into DSA in addition to the supra-thermal particles. The detailed physics for the injection process of these high-energy electrons coming from the ejecta is poorly understood and can be quite different from that for the supra-thermal electrons. Despite their high initial momenta compared to the supra-thermal electrons, not all of them reaching the FS can be accelerated when their incoming flux is high and their injection to DSA must be limited by the energy budget of their accelerator, i.e., the FS propagating into the CBM (see \S 3 for details). In this pilot study, for simplicity, we impose a conservative upper limit on their injection efficiency such that at any given time no more than 10\% of the incoming energy flux at the shock is used for their acceleration.\footnote{We note that with a high enough injection efficiency, the pressure from these accelerating electrons can feedback and change the shock structure as the protons do. We ignore this effect by only considering a moderate injection efficiency in this work. We expect to extend the work by implementing more detailed physical models in the future.}  

These accelerated electrons are then advected downstream from the FS while experiencing radiative and adiabatic energy losses within their respective Lagrangian fluid elements. In such way, space-integrated broadband synchrotron spectra are calculated as a function of time epoch to construct light curves in different energy bands. 

\section{Importance of the non-thermal electron injection by the radioactive decays}

Before showing the results of the simulations, we first provide an order-of-magnitude estimate on the importance of the contribution by the $\beta$-decay electrons to the non-thermal emission. For illustrative purpose, we show that this contribution can dominate the `traditional' supra-thermal electrons, assuming $M_\mrm{ej} = 0.01\ M_\odot$, $\widetilde{E_e} = 1$~MeV, no time delay, and $t \gg t_{\rm esc}$. The injection rate of the non-thermal electrons by the $\beta$-decays is 
\begin{equation}
N_{\rm e, decay} (t) \sim 6 \times 10^{42} \left(\frac{t}{1000\ {\rm days}}\right)^{-1.3} \ {\rm s}^{-1}  \ .
\end{equation}
On the other hand, the injection rate of the electrons through the shock heated supra-thermal particles is estimated as follows: 
\begin{equation}
N_{\rm e, sup-th} (t) \sim 4 \times 10^{40} \left(\frac{\eta_{\rm inj, th}}{10^{-4}}\right) \left(\frac{K_{\rm ep}}{0.01}\right)  \left(\frac{t}{1000\ {\rm days}}\right)^{2} \ {\rm s}^{-1} \ . 
\end{equation}
Therefore, the $\beta$-decay particles can dominate the injection for a few $1000$ days. 

Indeed, not all the $\beta$-decay electrons would be accelerated further by DSA from the energy budget consideration. {\it If all} the $\beta$-decay electrons would be accelerated to $\sim 100$ MeV, then the total energy injection by this processes would be $\sim 10^{39}$ erg s$^{-1}$ at $\sim 1,000$ days for our fiducial model parameters. On the other hand, the rate of the energy dissipation at the shock is roughly $\sim3 \times 10^{37}$ erg s$^{-1}$ at $\sim 1,000$ days (for $v_{\rm sk} \sim c$), assuming that the CBM is composed of hydrogen. This simple analysis shows that the electron acceleration should be limited by the available dissipated energy rather than the rate of the injected electrons. This situation is quite different than other astrophysical objects as the DSA site, and is a unique property of the NSM ejecta. This is why we set the upper limit for the electron acceleration by the available energy budget (\S 2). 

Another consideration to be made is on $t_\mrm{esc}$, the escape timescale of the $\beta$-decay electrons. \citet{2016ApJ...829..110B} compared two timescales; the thermalization timescale (i.e., the energy loss timescale assuming that the $\beta$-decay electrons are trapped within the ejecta), and the escape timescale  (i.e., the free-streaming timescale). The latter depends on the combination of the magnetic field configuration and the particle energy, i.e., a tangled magnetic field can `delay' the escaping time for a given electron energy. The thermalization timescale behaves as $\propto t^{-3}$, and thus the energy loss becomes quickly negligible. Therefore, for our situation, we can assume that the thermalization is negligible and the $\beta$-decay electrons are simply streaming outward freely in the ejecta. The escape timescale behaves as $\propto t$, the same as the dynamical (expansion) timescale, with the normalization given by the product of the magnetic configuration ($\lambda$ in \citet{2016ApJ...829..110B}) and the particle energy. As such, introducing a tangled magnetic field simply work as increasing the critical energy of the particle to catch up with the ejecta expansion. This may reduce the number of the $\beta$-decay particles reaching the FS, but indeed our acceleration is mostly limited by the kinetic energy dissipation at the FS, not by the number of $\beta$-decay particles at the FS (see \S 4). Therefore, our results are not much affected by the exact value of the escape timescale and therefore by the magnetic field configuration. Note that \citet{2016ApJ...829..110B} suggested the importance of the magnetic field configuration at the optical maximum by considering the competition between the thermalization timescale and escape timescale. This is however not important for our problem.

\section{Results and Discussion}

By constructing a number of hydrodynamical models, we explore the time evolution of radio and X-ray synchrotron emission behind the FS driven by the dynamical ejecta of a NSM event. We follow the evolution up to 1000 yrs after the ejection of the dynamical ejecta by the merger. Table~\ref{table1} summarizes the parameter space of our models. 

\begin{deluxetable*}{cccccc} 
\tablecolumns{6}
\tablewidth{10cm}
\tablecaption{Model Summary} 
\tablehead{\colhead{Model} & \colhead{$n_\mrm{CBM}$ (cm$^{-3}$)} & \colhead{$M_\mathrm{ej}$ ($M_\odot$)}  & \colhead{$E_\mrm{ej}$ ($10^{50}$ erg)} & \colhead{$\eta_\mrm{inj,th}$} & \colhead{$\beta$-decay $e^-$}   }
\startdata
1a		& $0.03$	& $0.04$	& $5.0$		& $3.3 \times 10^{-5}$	& No \\
1b		& $0.3$	& $0.04$	& $5.0$		& $3.3 \times 10^{-5}$	& No \\
1c		& $0.03$	& $0.01$	& $1.25$		& $3.3 \times 10^{-5}$	& No \\
1d		& $0.03$	& $0.04$	& $5.0$		& $4.2 \times 10^{-4}$	& No \\
2a		& $0.03$	& $0.04$	& $5.0$		& $3.3 \times 10^{-5}$	& Yes \\
2b		& $0.3$	& $0.04$	& $5.0$		& $3.3 \times 10^{-5}$	& Yes \\
2c		& $0.03$	& $0.01$	& $1.25$		& $3.3 \times 10^{-5}$	& Yes \\
2d		& $0.03$	& $0.04$	& $5.0$		& $4.2 \times 10^{-4}$	& Yes 
\enddata
\label{table1}  
\end{deluxetable*}

The models are labeled in the following manner. Models without and with the acceleration of $\beta$-decay electrons injected by the ejecta are labeled `1' and `2' respectively. For each of these two groups, the models `a' are considered as our fiducial models, which are compared with the models `b' with a 10-fold higher circumbinary gas density ($n_\mrm{CBM}$), and the models `c' with a 4 times lower initial ejecta kinetic energy ($E_\mrm{ej}$) and mass ($M_\mrm{ej}$). We further add the models `d' to investigate the effect of an enhanced DSA injection efficiency ($\eta_\mrm{inj,th}$) for the post-shock supra-thermal particles (of which the microphysics is still not completely understood) on the behavior of the radio and X-ray light curves. We assume a distance of 40~Mpc for the NSM event in question, and a circumbinary magnetic field strength $B_0 = 3\ \mu$G.

\subsection{Model dependence}

\begin{figure}
\centering
\includegraphics[width=8.5cm]{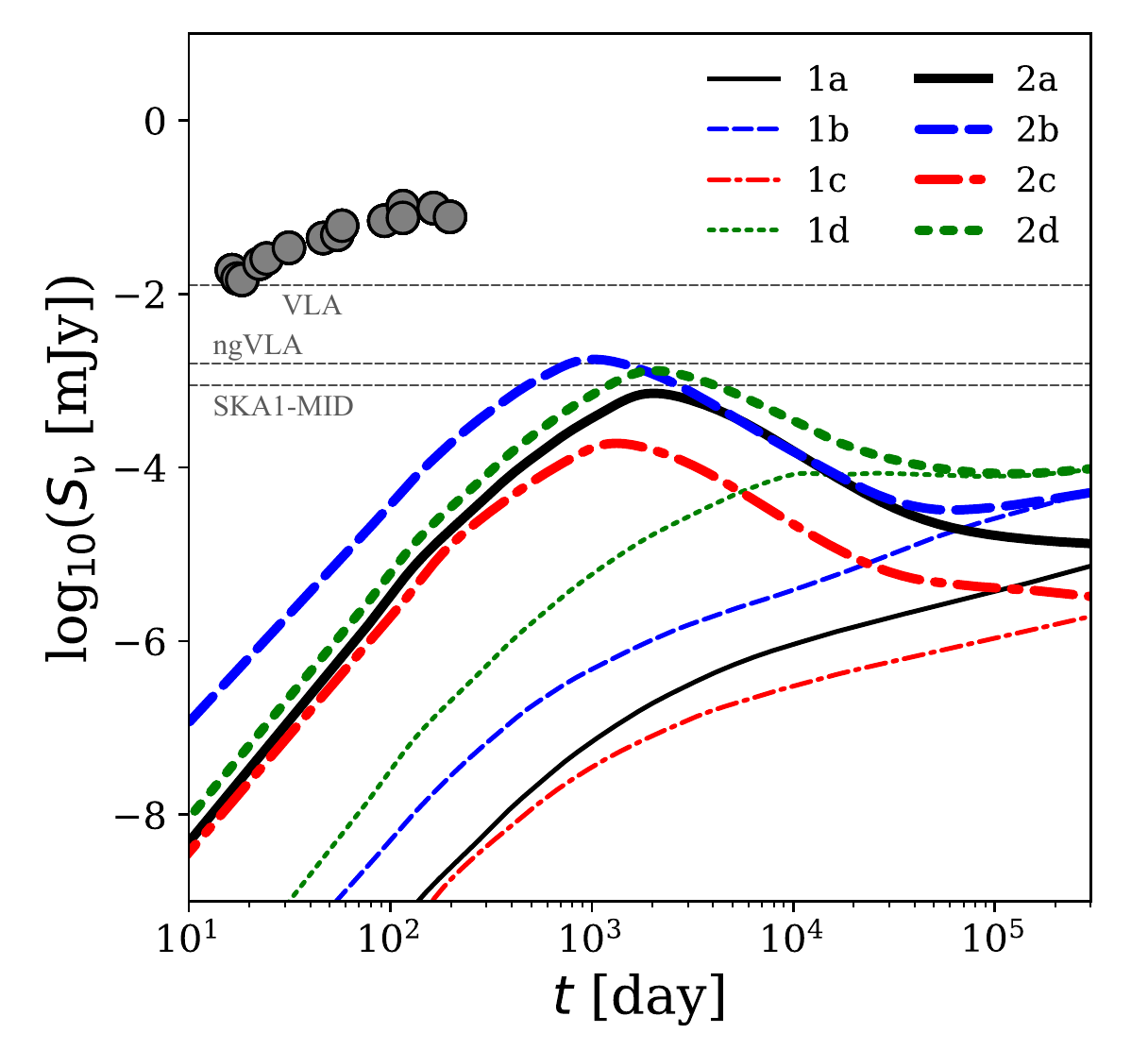}  
\caption{
3 GHz radio light curve up to 1000 yrs predicted by our models listed in Table~\ref{table1}. Models 1a to 1d do not account for the acceleration of $\beta$-decay electrons while models 2a to 2c do. Our fiducial models are marked by label `a', which are compared to models with a higher circumbinary gas density (label `b'), a lower explosion energy and ejecta mass (label `c'), and a higher DSA injection efficiency for the thermally injected electrons (label `d'). Our models show that the contribution of the accelerated $\beta$-decay electrons dominate the earlier radio continuum emission over the thermally injected electrons up to a few 100 yrs. For comparison, observational data for the NSM event GW170817 from \textit{ATCA} and \textit{VLA} are overlaid as grey circles \citep{Hallinan1579,2018Natur.554..207M,2018arXiv180103531M,2018arXiv180306853D}. The thin grey dashed lines show the sensitivities of current and future instruments for an 1 hr exposure. 
}
\label{fig3}
\end{figure}

Figure~\ref{fig3} shows the light curves for the 3~GHz radio continuum flux density predicted by our models. Without the contribution from the $\beta$-decay electrons, the fiducial model 1a shows a typical rising light curve as the FS expands into the uniform CBM and emission volume increases. The light curve flattens gradually as the FS decelerates with time, and the accelerated electrons suffer from adiabatic loss. At a moderate DSA injection level $\eta_\mrm{inj,th} \sim 3 \times 10^{-5}$, low circumbinary gas density $n_\mrm{CBM} = 0.03$ cm$^{-3}$ and downstream magnetic field strength $B \sim 30\ \mu$G, the flux density is $S_\nu \sim 10^{-5}$~mJy after 1000~yrs, and continues to rise gradually as the ejecta  expands into the low-density circumbinary medium. The flux density eventually start to decline with time as $t^{-3/2}$ at $\sim 10^4$~yr after merger as the FS decelerates further.

The effect of a higher CBM density on the radio emission is demonstrated by comparing model 1b to 1a. Here, $n_\mrm{CBM}$ is increased by a factor of 10 to $0.3$ cm$^{-3}$. The injection rate of the thermal particles into DSA is hence 10 times higher in model 1b. The downstream magnetic field strength is boosted to $50-60\ \mu$G due to a more efficient MFA in the shock precursor, which is about a factor of 2 higher than model 1a on average. On the other hand, the emission volume at a given time is smaller than model 1a since the FS also decelerates faster with time in the  denser environment. At 1000 yrs, the shock radius is 5.4 pc in model 1b, compared to 8.7 pc in model 1a. Since the radio flux density is proportional to $K_{ep}\eta_\mrm{inj,th}n_\mrm{CBM}BV$ where $V$ is the emission volume, the result shows an enhancement of the flux by a factor $\sim$ 5 to 10 throughout the simulation time as expected. 

In the case of the less massive dynamical ejecta (model 1c), the FS decelerates with time faster as in model 1b. 
The DSA injection rate per volume ($\propto K_{ep}\eta_\mrm{inj,th}n_\mrm{CBM}R_\mrm{tot}$) however is similar to the fiducial model as the shock compression ratio $R_\mrm{tot}$ ranges between $4.0$ to $4.3$ for both models.
At 1000 yrs, the FS radius is 6.6 pc, and the magnetic field strength is similar to model 1a. The result is hence a radio light curve that rises more gradually. The reduction of the ejecta mass by a factor of 4 leads to a weaker radio emission in overall by roughly the same factor. 

Model 1d shows the response of the light curve to an enhanced DSA injection efficiency for the thermally injected particles. The injection number fraction $\eta_\mrm{inj,th}$ is increased by a factor $\sim 10$. In addition to a boost of the number of radio-emitting electrons by the same factor, it also results in a more effective MFA in the shock precursor. The downstream $B$-field strength is roughly $100$ -- $340\ \mu$G in model 1d. These effects contribute to the significant brightening of the radio emission compared to the fiducial model as shown in Figure~\ref{fig3}. Model 2d adds the early contribution from the $\beta$-decay electrons. The radio emission is overall brighter than the corresponding fiducial model 2a due to the higher $B$-field, and converges back with model 1d after $\sim 100$~yrs.     

By comparing the fiducial models 2a and 1a, we see that the injection of $\beta$-decay electrons, whose luminosity peaks at around 10 days (see Figure~\ref{fig2}), enhances the early-phase radio emission substantially up to $\sim 10^4$ days. This confirms our order-of-magnitude estimate in \S 3. Indeed, the injection of these $\beta$-decay electrons is `too efficient', and the acceleration process in such a regime has not been clarified. From the energy budget consideration, we simply set the upper limit so that 10\% of the incoming energy flux at the shock is used to accelerate the electrons, as mentioned in \S 2. While the quantitative behavior may depend on the treatment of the acceleration process in this highly efficient regime, our simulations illustrate the importance of the $\beta$-decay electrons in shaping the non-thermal emission from the NSM dynamical ejecta.

These electrons dominate the emission in the first few 100 yrs until the flux of the $\beta$-decay electrons from the ejecta has decreased enough at the FS location and the accelerated $\beta$-decay electrons have cooled off by adiabatic expansion in the post-shock region. After that, the radio flux is dominated by the thermally injected electrons so that the light curve converges back onto model 1a. The 3~GHz flux density peaks at $\sim 10^{-3}$~mJy, roughly 4 order-of-magnitudes higher than model 1a at the same evolutionary stage, and the peak occurs in a much earlier phase at around 2000 days. 

The reactions of the radio light curve to a higher CBM density is exhibited by model 2b. The higher $n_\mrm{CBM}$ in model 2b results in a higher FS ram pressure, i.e., $P_\mrm{sk} = \rho_\mrm{CBM}v_\mrm{sk}^2/2$. Since the injection of the $\beta$-decay electrons in the early phase ($\sim 1$~yr) is dictated by the upper limit we imposed (see above), the increase in $P_\mrm{sk}$ also means a corresponding increase in the number of electrons being accelerated. The resulting radio flux is therefore also roughly an order-of-magnitude higher for model 2b than 2a. After that, the bolometric luminosity of these electrons has dropped with time significantly so that the above upper limit is no longer effective. The electrons injection rate therefore becomes identical between the two models, and the light curves become similar in the rest of the phase in which the $\beta$-decay electrons still dominate the emission. Other effects of a denser environment on the evolution of emission volume and magnetic field strength as mentioned above still exist.

The effect of a different ejecta mass is covered by model 2c. Model 2a and 2c show similar light curves in the rising phase up to $\sim 1$~yr. This again reflects the upper limit we impose on the number of accelerated $\beta$-decay electrons by the energy budget available at the FS. The difference in the ejecta mass, and therefore the number of seed electrons emitted by the ejecta, is generally hidden in the rising part of the light curve. As the bolometric luminosity of these electrons has dropped with time, however, its time dependence and proportionality to $M_\mrm{ej}$ start to take effect on the radio emission, so that the radio flux density of model 2c becomes systematically smaller than model 2a after $\sim 1$~yr. The smaller emission volume for model 2c also plays a role.        

An enhanced injection efficiency of the thermal particles (model 2d) does not show noticeable effect on the light curve in the first $\sim 1000$ days, which is as expected because the relativistic electrons in this phase are dominated by the $\beta$-decay electrons. Like model 2a, the light curves of model 2b, 2c and 2d eventually converge back with model 1b, 1c and 1d respectively as the thermally injected electrons start to dominate the emission. 

%
\begin{figure}
\centering
\includegraphics[width=8.5cm]{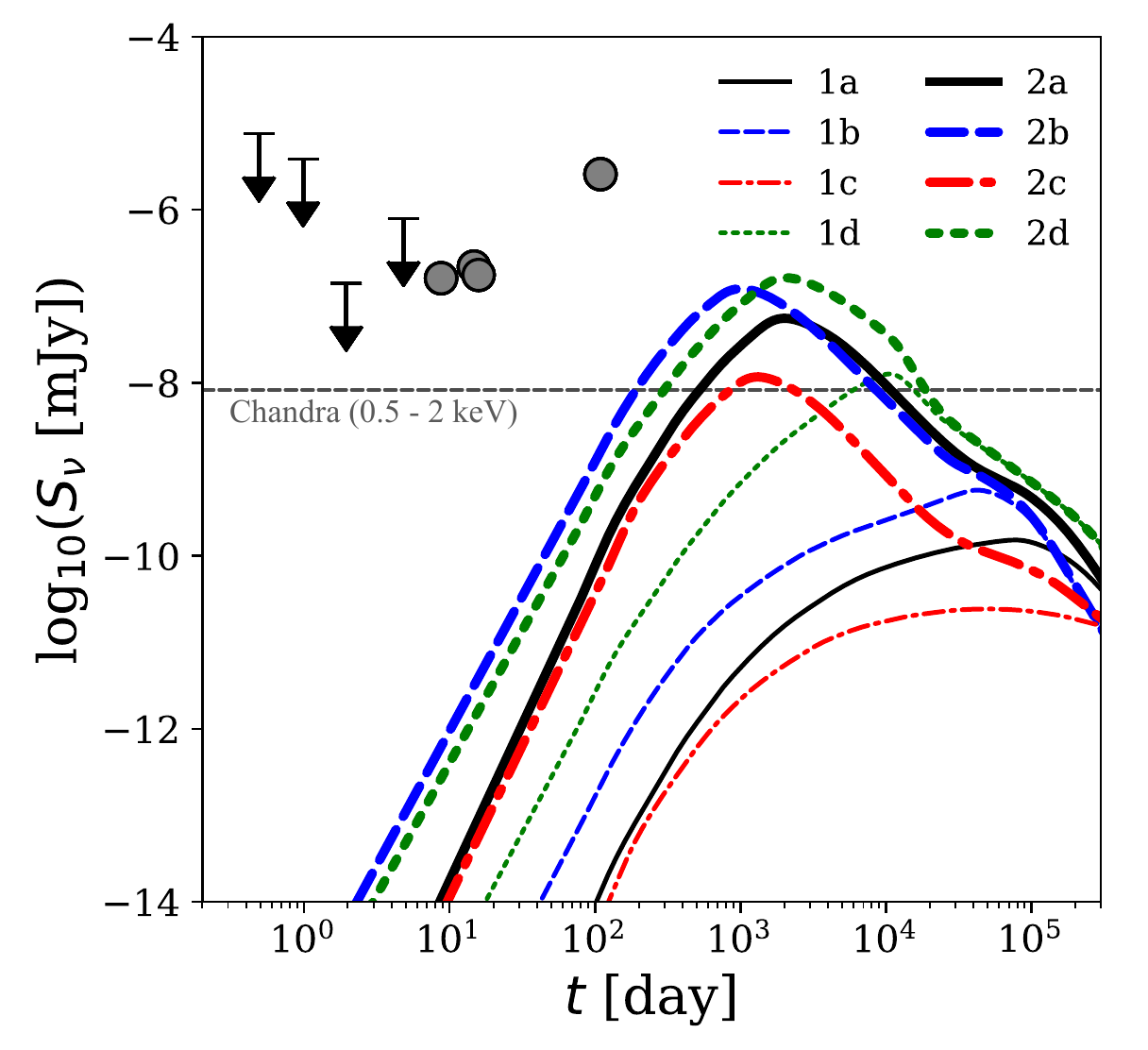}  
\caption{
Same as Figure~\ref{fig3} but for synchrotron X-rays at 1~keV. For comparison, \textit{Chandra} data for the NSM event GW170817 are overlaid as upper-limits and circles \citep{2017Natur.551...71T}. The thin grey dashed line shows the sensitivity of \textit{Chandra} in the 0.5 -- 2 keV band for a 1 Ms exposure.
}
\label{fig4}
\end{figure}

Very similar behaviors can be observed in the X-ray light curves (see Figure~\ref{fig4}). The main difference between the 1 keV and 3 GHz emission is that the former is sensitive to the maximum energy $E_\mrm{max,e}(t)$ of the accelerated electrons, and hence the synchrotron cutoff frequency $\nu_\mrm{cut}(t) \propto E^2_\mrm{max,e} B$, which is limited by both the acceleration and synchrotron loss timescales. A break feature appears in the light curves when $E_\mrm{max,e}$ has decreased due to synchrotron loss to a point such that $h\nu_\mrm{cut} \lesssim 1$~keV. As a result, models with higher downstream $B$-field, e.g., model 1d, have such a break appearing earlier on in the light curve. 

In the early phase of the evolution, we also observe a higher X-ray flux from model 2d than the fiducial model 2a. This is again mainly attributed to a stronger $B$-field amplified by a more efficient DSA of protons at the FS. Initially, $E_\mrm{max,e}$ is limited by the acceleration timescale which is inversely proportional to $B$. A shorter acceleration timescale and hence higher $E_\mrm{max,e}$ results in a higher $h\nu_\mrm{cut}$, which at $\sim$ 100 days is $\sim$ 10~keV for model 2d compared to $\sim$ 0.8~keV for model 2a. The brightness at 1~keV is hence less affected by the spectral cutoff for model 2d in the early phase. After a few yrs, $E_\mrm{max,e}$ becomes limited by synchrotron loss just like model 1d.   

In both Figure~\ref{fig3} and \ref{fig4}, the recently observed radio and X-ray data from the GW170817 event are displayed. According to our models, it is obvious that shock acceleration triggered by the dynamical ejecta does not contribute importantly to the emission in the early phase. It has been proposed that an off-axis jet \citep[e.g.,][]{2017arXiv171203237L} can satisfactorily explain the broadband data for this particular event (see \citet{2018arXiv180300599H} for an alternative interpretation). Therefore, it is instructive to make a comparison of our prediction with such models for the jet component in order to discuss the future detectability of the emission from the dynamical ejecta, especially for other NSM events to be detected in the future.

\subsection{Comparison with jet component and detectability}

\begin{figure}
\centering
\includegraphics[width=8.5cm]{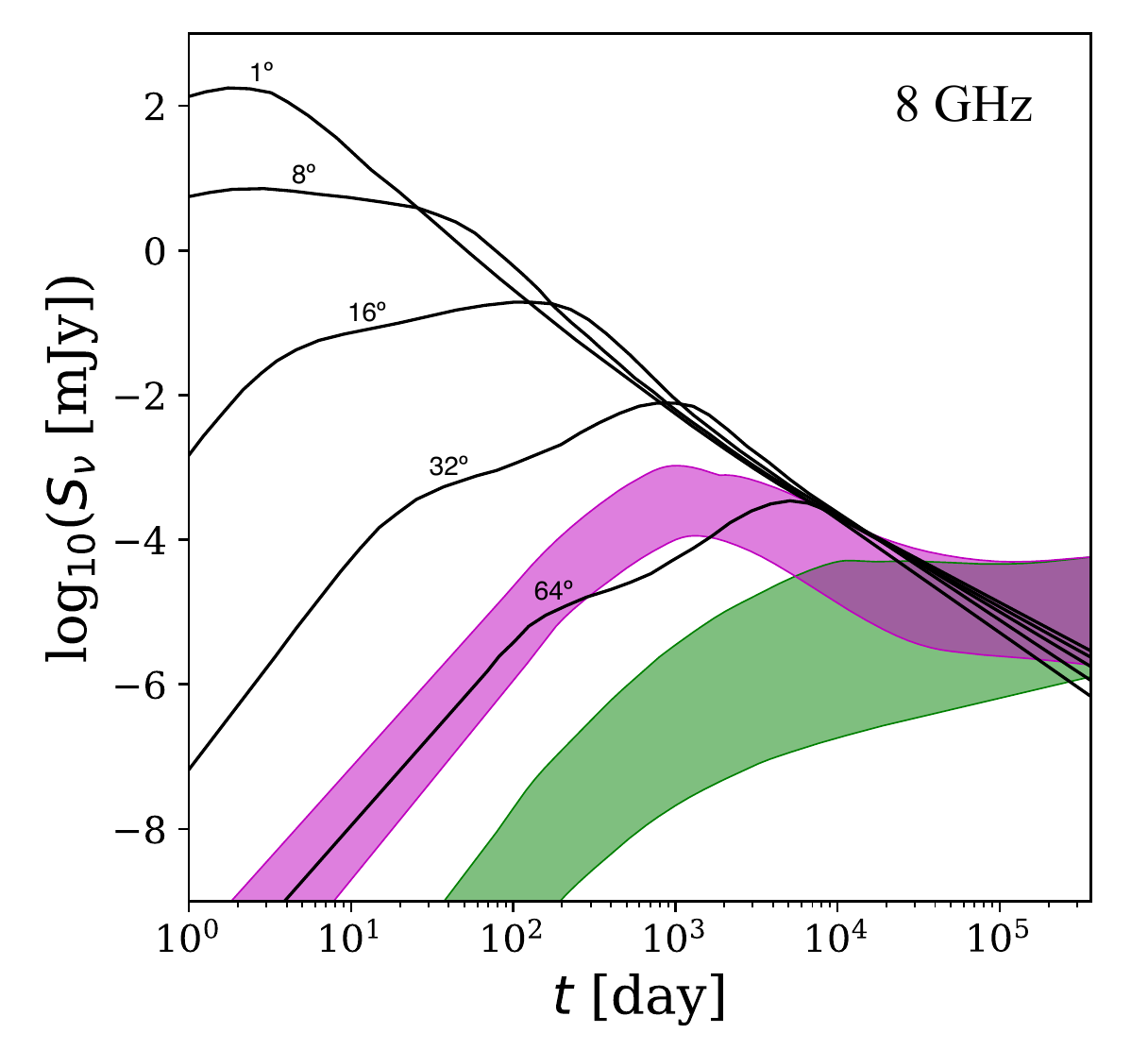}  
\caption{
Comparison of 8 GHz radio light curves of a structured jet model at different viewing angles against the jet-axis by \citet{2017arXiv171203237L} (thin black lines) with our models for shock emission produced by dynamical ejecta. The range of radio flux densities spanned by the parameter space of our models with and without acceleration of $\beta$-decay electrons are shown by the magenta and green colored bands respectively.
} 
\label{fig5}
\end{figure}

\begin{figure}
\centering
\includegraphics[width=8.5cm]{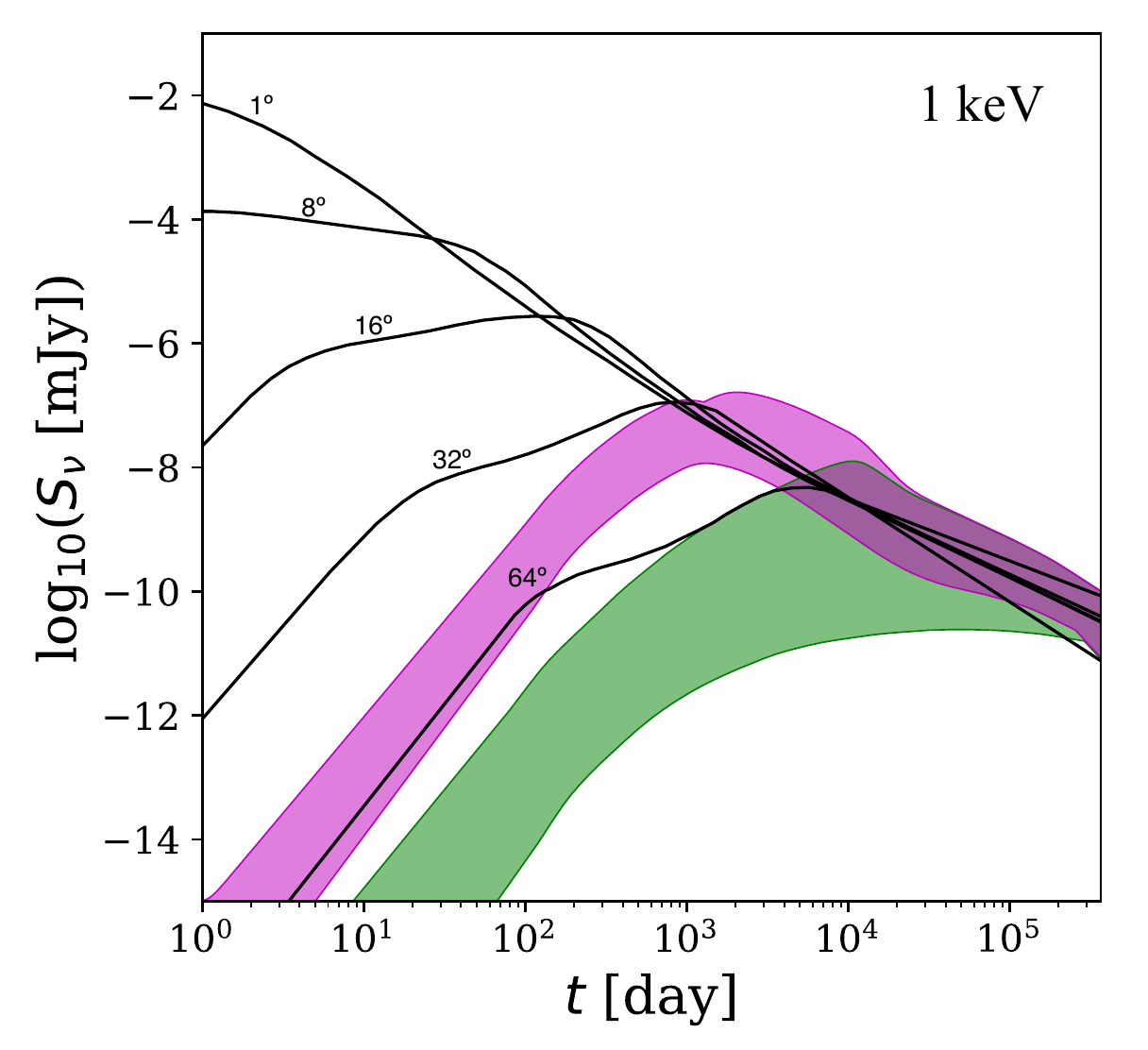}  
\caption{
Same as Figure~\ref{fig5} but for synchrotron X-rays at 1 keV.} 
\label{fig6}
\end{figure}

To assess the observability of this synchrotron emission originating from the dynamical ejecta, it is meaningful to compare our light curves with those predicted for the jet component which has been suggested to be responsible for the short GRB associated with a NSM event. In Figure~\ref{fig5} and \ref{fig6}, we overlay light curves from the models by \citet{2017arXiv171203237L} for an off-axis jet on our models. 
The parameters for these jet models are tuned to match the observed light curves from the GW170817 event. The off-axis jet models for several observer viewing angles $\theta$ against the jet-axis are shown. As reference, the case for $\theta = 16^\circ$ is close to the best-fit model for GW170817. The colored bands cover the range of the predicted light curves from our models for the dynamical ejecta component with (magenta) and without (green) the contribution from accelerated $\beta$-decay electrons. 

In the near future, the number of detected nearby NSMs is expected to increase, and then they should show variation depending on the viewing direction. Therefore, it is instructive to see if the contribution from the dynamical ejecta in the non-thermal emission is visible beyond the other component(s). We can see that the jet component dominates the broadband emission for the first few 100 yrs if the jet-axis makes an angle with the line-of-sight shallower than $\sim 50^\circ$. For cases with the accelerated $\beta$-decay electrons included, the dynamical ejecta component becomes comparable to the jet component if an observer is placed at the angle $> 50^\circ$. At a distance of 40~Mpc and energetics similar to that inferred from GW170817, the peak flux density occurs at about 1000 days, at a level of about $10^{-4}$ -- $10^{-3}$~mJy (8~GHz) and $10^{-8}$ -- $10^{-7}$~mJy (1~keV). 

%

Figure~\ref{fig3} and \ref{fig7} compare our radio light curves at 3 and 6~GHz respectively with the sensitivities of several instruments in the corresponding observation wavebands, including the \textit{Very Large Array (VLA)}, and the upcoming \textit{next-generation Very Large Array (ngVLA)} and \textit{Square Kilometre Array (SKA)}. The model by \citet{2017ApJ...848L..21A} at 6~GHz is also shown in Figure~\ref{fig7} for comparison (see \S 5 for comparison between their models and our models). Our optimistic models predict that the radio emission from ejecta-CBM interaction in a NSM event at $40$~Mpc achieves a peak flux density comparable to the 1-hr sensitivity of next-generation instruments in the near future.

The late-time radio emission in our models is dominated by the thermally injected electrons. In the optimistic cases, the 3 GHz flux density is about $10^{-4}$~mJy at 40 Mpc, and remains nearly constant until $\sim 10^4$~yr after the merger. Since the binary neutron star merger rate in the Milky way is estimated from known Galactic binary neutron star systems as $R_\mrm{MW} = 21 \pm^{28}_{14}$~Myr$^{-1}$ \citep{2015MNRAS.448..928K}, the expected number of such radio sources in the Milky Way is only $\sim$ 0.2, which does not give any constraint on these models. As for nearby galaxies, the BNS merger rate density is estimated as $\sim 1540 \pm^{3200}_{1220}$~Gpc$^{-3}$yr$^{-1}$ from the detection of GW170817 \citep{2017PhRvL.119p1101A}, and then the expectation number of radio sources detectable by the all-sky survey by \textit{VLA} (\textit{VLASS}) at 3 GHz with a combined sensitivity of $69\ \mu$Jy is $\sim$ 0.05 -- 0.7. Using \textit{ngVLA} and \textit{SKA}, whose sensitivities are about 10 times better than \textit{VLA} at 3 GHz, the expectation number increases to $\sim$ 1.6 -- 22. However, since the BNS merger rate density in the local universe is still not well constrained from the GW observations, we still cannot give any constraint on our model from these estimates.  On the other hand, the X-ray emission around the peak is well above the sensitivity of \textit{Chandra} for the optimistic cases.  Assuming a 1 Ms exposure, the detectable distance of \textit{Chandra} would be $\sim$ 130 Mpc.  We may verify or rule out our model by searching for this X-ray emission associating with the GW emission from NSMs.


\section{Concluding Remarks}
%
\begin{figure}
\centering
\includegraphics[width=8.5cm]{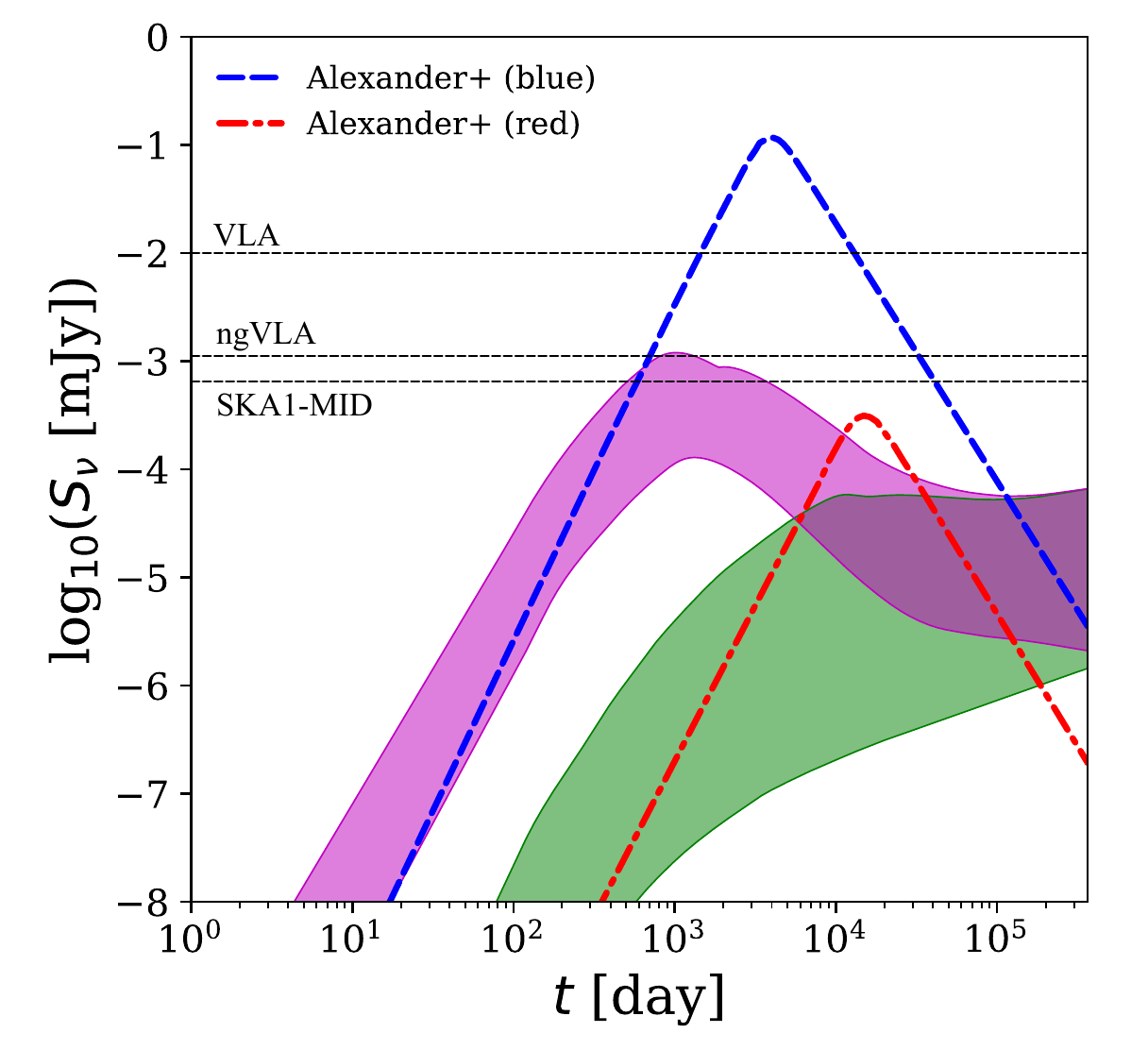}  
\caption{
Comparison of 6~GHz light curves from our models with those by \citet{2017ApJ...848L..21A} for the blue kilonova component (blue dashed line) and the red kilonova component (red dash-dotted line) scaled to $n_\mrm{CBM} = 0.03$~cm$^{-3}$. The thin dashed lines again show the sensitivities of current and future instruments for an 1 hr exposure.} 
\label{fig7}
\end{figure}

In this study, through the construction of hydrodynamical models and parametric survey, we show that NSM events can shine in radio and X-ray with significant contribution from the shock acceleration of $\beta$-decay electrons injected by the expanding dynamical ejecta. While the emission is relatively weak compared to the accompanying jet component, with a larger viewing angle against the jet-axis and a relatively close distance, it is possible that such emission can be detected in the near future by next-generation telescopes. Furthermore, it is possible that some NSM events do not launch a jet (so-called a `choked jet'), and in such cases we may observe the emission from $\beta$-decay electrons as a dominant component. 

Our simulations without the $\beta$-decay electron injection result in much fainter radio emission than previously predicted with semi-analytic calculations with a parameterized treatment of the electron acceleration and magnetic field amplification \citep[e.g.,][]{2016ApJ...831..190H,2017ApJ...848L..21A,2018arXiv180300599H}. Figure~\ref{fig7} shows that the difference reaches to roughly 3 to 5 orders of magnitude in the predicted radio flux. This stems from the much lower efficiency of the electron acceleration than optimistically assumed in these previous calculations, once we adopt the physically-motivated recipe to the acceleration and magnetic field amplification (as calibrated by the non-thermal emission from SNRs). Indeed, with the $\beta$-decay electrons, we recover the radio flux similar to the previous models in the rising part (up to $\sim 1000$ days). With the $\beta$-decay electrons, the efficiency of the acceleration can reach to nearly equipartition to the shock dissipated energy, and thus the situation becomes similar to the previous studies. It is hard to realize such high efficiency in the canonical DSA of thermal electrons, highlighting the importance of the seed electrons from the $\beta$-decays within the NSM ejecta. 

Our work highlights a unique nature of the non-thermal emission from the NSM ejecta. This site represents a regime in which the seed electrons for DSA acceleration are extremely abundant, a situation which has never been known to be realized in other astrophysical DSA sites. 
We anticipate that \textit{Advanced LIGO} and \textit{VIRGO} and other upcoming gravitational wave detectors such as \textit{KAGRA} will pick up more and more GW signals from NSM events in the local Universe, and the sample of their associated electromagnetic counterparts will enlarge. The possible detection of synchrotron emission from ejecta-CBM interaction at NSMs, and the shock acceleration of $\beta$-decay electrons, will have profound implications including r-process nucleosynthesis process at NSM events and the origin of neutron-rich heavy elements in the Universe.


\acknowledgments 
KM acknowledges support from KAKENHI Grant 17H02864. NK acknowledges support from the Hakubi Project at Kyoto University.

\bibliographystyle{aa} 
\bibliography{reference}

\begin{thebibliography}{71}
\expandafter\ifx\csname natexlab\endcsname\relax\def\natexlab#1{#1}\fi

\bibitem[{{Abbott} {et~al.}(2017{\natexlab{a}}){Abbott}, {Abbott}, {Abbott},
  {Acernese}, {Ackley}, {Adams}, {Adams}, {Addesso}, {Adhikari}, {Adya}, \&
  et~al.}]{2017ApJ...848L..13A}
{Abbott}, B.~P., {Abbott}, R., {Abbott}, T.~D., {et~al.} 2017{\natexlab{a}},
  \apjl, 848, L13

\bibitem[{{Abbott} {et~al.}(2017{\natexlab{b}}){Abbott}, {Abbott}, {Abbott},
  {Acernese}, {Ackley}, {Adams}, {Adams}, {Addesso}, {Adhikari}, {Adya}, \&
  et~al.}]{2017PhRvL.119p1101A}
{Abbott}, B.~P., {Abbott}, R., {Abbott}, T.~D., {et~al.} 2017{\natexlab{b}},
  Physical Review Letters, 119, 161101

\bibitem[{{Abbott} {et~al.}(2017{\natexlab{c}}){Abbott}, {Abbott}, {Abbott},
  {Acernese}, {Ackley}, {Adams}, {Adams}, {Addesso}, {Adhikari}, {Adya}, \&
  et~al.}]{2017ApJ...848L..12A}
{Abbott}, B.~P., {Abbott}, R., {Abbott}, T.~D., {et~al.} 2017{\natexlab{c}},
  \apjl, 848, L12

\bibitem[{{Alexander} {et~al.}(2017){Alexander}, {Berger}, {Fong}, {Williams},
  {Guidorzi}, {Margutti}, {Metzger}, {Annis}, {Blanchard}, {Brout}, {Brown},
  {Chen}, {Chornock}, {Cowperthwaite}, {Drout}, {Eftekhari}, {Frieman}, {Holz},
  {Nicholl}, {Rest}, {Sako}, {Soares-Santos}, \&
  {Villar}}]{2017ApJ...848L..21A}
{Alexander}, K.~D., {Berger}, E., {Fong}, W., {et~al.} 2017, \apjl, 848, L21

\bibitem[{{Andreoni} {et~al.}(2017){Andreoni}, {Ackley}, {Cooke}, {Acharyya},
  {Allison}, {Anderson}, {Ashley}, {Baade}, {Bailes}, {Bannister}, {Beardsley},
  {Bessell}, {Bian}, {Bland}, {Boer}, {Booler}, {Brandeker}, {Brown},
  {Buckley}, {Chang}, {Coward}, {Crawford}, {Crisp}, {Crosse}, {Cucchiara},
  {Cup{\'a}k}, {de Gois}, {Deller}, {Devillepoix}, {Dobie}, {Elmer}, {Emrich},
  {Farah}, {Farrell}, {Franzen}, {Gaensler}, {Galloway}, {Gendre}, {Giblin},
  {Goobar}, {Green}, {Hancock}, {Hartig}, {Howell}, {Horsley}, {Hotan},
  {Howie}, {Hu}, {Hu}, {James}, {Johnston}, {Johnston-Hollitt}, {Kaplan},
  {Kasliwal}, {Keane}, {Kenney}, {Klotz}, {Lau}, {Laugier}, {Lenc}, {Li},
  {Liang}, {Lidman}, {Luvaul}, {Lynch}, {Ma}, {Macpherson}, {Mao},
  {McClelland}, {McCully}, {M{\"o}ller}, {Morales}, {Morris}, {Murphy},
  {Noysena}, {Onken}, {Orange}, {Os{\l}owski}, {Pallot}, {Paxman}, {Potter},
  {Pritchard}, {Raja}, {Ridden-Harper}, {Romero-Colmenero}, {Sadler}, {Sansom},
  {Scalzo}, {Schmidt}, {Scott}, {Seghouani}, {Shang}, {Shannon}, {Shao},
  {Shara}, {Sharp}, {Sokolowski}, {Sollerman}, {Staff}, {Steele}, {Sun},
  {Suntzeff}, {Tao}, {Tingay}, {Towner}, {Thierry}, {Trott}, {Tucker},
  {V{\"a}is{\"a}nen}, {Krishnan}, {Walker}, {Wang}, {Wang}, {Wayth}, {Whiting},
  {Williams}, {Williams}, {Wolf}, {Wu}, {Wu}, {Yang}, {Yuan}, {Zhang}, {Zhou},
  \& {Zovaro}}]{2017PASA...34...69A}
{Andreoni}, I., {Ackley}, K., {Cooke}, J., {et~al.} 2017, PASA, 34, e069

\bibitem[{{Arcavi} {et~al.}(2017){Arcavi}, {Hosseinzadeh}, {Howell}, {McCully},
  {Poznanski}, {Kasen}, {Barnes}, {Zaltzman}, {Vasylyev}, {Maoz}, \&
  {Valenti}}]{2017Natur.551...64A}
{Arcavi}, I., {Hosseinzadeh}, G., {Howell}, D.~A., {et~al.} 2017, \nat, 551, 64

\bibitem[{{Asano} \& {To}(2018)}]{2018ApJ...852..105A}
{Asano}, K. \& {To}, S. 2018, \apj, 852, 105

\bibitem[{{Barnes} \& {Kasen}(2013)}]{2013ApJ...775...18B}
{Barnes}, J. \& {Kasen}, D. 2013, \apj, 775, 18

\bibitem[{{Barnes} {et~al.}(2016){Barnes}, {Kasen}, {Wu}, \&
  {Mart{\'{\i}}nez-Pinedo}}]{2016ApJ...829..110B}
{Barnes}, J., {Kasen}, D., {Wu}, M.-R., \& {Mart{\'{\i}}nez-Pinedo}, G. 2016,
  \apj, 829, 110

\bibitem[{{Blasi} {et~al.}(2005){Blasi}, {Gabici}, \& {Vannoni}}]{BGV2005}
{Blasi}, P., {Gabici}, S., \& {Vannoni}, G. 2005, \mnras, 361, 907

\bibitem[{{Bromberg} {et~al.}(2017){Bromberg}, {Tchekhovskoy}, {Gottlieb},
  {Nakar}, \& {Piran}}]{2017arXiv171005897B}
{Bromberg}, O., {Tchekhovskoy}, A., {Gottlieb}, O., {Nakar}, E., \& {Piran}, T.
  2017, ArXiv e-prints

\bibitem[{{Caprioli} {et~al.}(2009){Caprioli}, {Blasi}, {Amato}, \&
  {Vietri}}]{CBAV2009}
{Caprioli}, D., {Blasi}, P., {Amato}, E., \& {Vietri}, M. 2009, \mnras, 395,
  895

\bibitem[{{Chornock} {et~al.}(2017){Chornock}, {Berger}, {Kasen},
  {Cowperthwaite}, {Nicholl}, {Villar}, {Alexander}, {Blanchard}, {Eftekhari},
  {Fong}, {Margutti}, {Williams}, {Annis}, {Brout}, {Brown}, {Chen}, {Drout},
  {Farr}, {Foley}, {Frieman}, {Fryer}, {Herner}, {Holz}, {Kessler}, {Matheson},
  {Metzger}, {Quataert}, {Rest}, {Sako}, {Scolnic}, {Smith}, \&
  {Soares-Santos}}]{2017ApJ...848L..19C}
{Chornock}, R., {Berger}, E., {Kasen}, D., {et~al.} 2017, \apjl, 848, L19

\bibitem[{Coulter {et~al.}(2017)Coulter, Foley, Kilpatrick, Drout, Piro,
  Shappee, Siebert, Simon, Ulloa, Kasen, Madore, Murguia-Berthier, Pan,
  Prochaska, Ramirez-Ruiz, Rest, \& Rojas-Bravo}]{Coulter1556}
Coulter, D.~A., Foley, R.~J., Kilpatrick, C.~D., {et~al.} 2017, Science, 358,
  1556

\bibitem[{{Cowperthwaite} {et~al.}(2017){Cowperthwaite}, {Berger}, {Villar},
  {Metzger}, {Nicholl}, {Chornock}, {Blanchard}, {Fong}, {Margutti},
  {Soares-Santos}, {Alexander}, {Allam}, {Annis}, {Brout}, {Brown}, {Butler},
  {Chen}, {Diehl}, {Doctor}, {Drout}, {Eftekhari}, {Farr}, {Finley}, {Foley},
  {Frieman}, {Fryer}, {Garc{\'{\i}}a-Bellido}, {Gill}, {Guillochon}, {Herner},
  {Holz}, {Kasen}, {Kessler}, {Marriner}, {Matheson}, {Neilsen}, {Quataert},
  {Palmese}, {Rest}, {Sako}, {Scolnic}, {Smith}, {Tucker}, {Williams},
  {Balbinot}, {Carlin}, {Cook}, {Durret}, {Li}, {Lopes}, {Louren{\c c}o},
  {Marshall}, {Medina}, {Muir}, {Mu{\~n}oz}, {Sauseda}, {Schlegel}, {Secco},
  {Vivas}, {Wester}, {Zenteno}, {Zhang}, {Abbott}, {Banerji}, {Bechtol},
  {Benoit-L{\'e}vy}, {Bertin}, {Buckley-Geer}, {Burke}, {Capozzi}, {Carnero
  Rosell}, {Carrasco Kind}, {Castander}, {Crocce}, {Cunha}, {D'Andrea}, {da
  Costa}, {Davis}, {DePoy}, {Desai}, {Dietrich}, {Drlica-Wagner}, {Eifler},
  {Evrard}, {Fernandez}, {Flaugher}, {Fosalba}, {Gaztanaga}, {Gerdes},
  {Giannantonio}, {Goldstein}, {Gruen}, {Gruendl}, {Gutierrez}, {Honscheid},
  {Jain}, {James}, {Jeltema}, {Johnson}, {Johnson}, {Kent}, {Krause}, {Kron},
  {Kuehn}, {Nuropatkin}, {Lahav}, {Lima}, {Lin}, {Maia}, {March}, {Martini},
  {McMahon}, {Menanteau}, {Miller}, {Miquel}, {Mohr}, {Neilsen}, {Nichol},
  {Ogando}, {Plazas}, {Roe}, {Romer}, {Roodman}, {Rykoff}, {Sanchez},
  {Scarpine}, {Schindler}, {Schubnell}, {Sevilla-Noarbe}, {Smith}, {Smith},
  {Sobreira}, {Suchyta}, {Swanson}, {Tarle}, {Thomas}, {Thomas}, {Troxel},
  {Vikram}, {Walker}, {Wechsler}, {Weller}, {Yanny}, \&
  {Zuntz}}]{2017ApJ...848L..17C}
{Cowperthwaite}, P.~S., {Berger}, E., {Villar}, V.~A., {et~al.} 2017, \apjl,
  848, L17

\bibitem[{{Dobie} {et~al.}(2018){Dobie}, {Kaplan}, {Murphy}, {Lenc}, {Mooley},
  {Lynch}, {Corsi}, {Frail}, {Kasliwal}, \& {Hallinan}}]{2018arXiv180306853D}
{Dobie}, D., {Kaplan}, D.~L., {Murphy}, T., {et~al.} 2018, ArXiv e-prints

\bibitem[{Drout {et~al.}(2017)Drout, Piro, Shappee, Kilpatrick, Simon,
  Contreras, Coulter, Foley, Siebert, Morrell, Boutsia, Di~Mille, Holoien,
  Kasen, Kollmeier, Madore, Monson, Murguia-Berthier, Pan, Prochaska,
  Ramirez-Ruiz, Rest, Adams, Alatalo, Ba{\~n}ados, Baughman, Beers, Bernstein,
  Bitsakis, Campillay, Hansen, Higgs, Ji, Maravelias, Marshall, Bidin, Prieto,
  Rasmussen, Rojas-Bravo, Strom, Ulloa, Vargas-Gonz{\'a}lez, Wan, \&
  Whitten}]{Drout1570}
Drout, M.~R., Piro, A.~L., Shappee, B.~J., {et~al.} 2017, Science, 358, 1570

\bibitem[{{Ellison} {et~al.}(2007){Ellison}, {Patnaude}, {Slane}, {Blasi}, \&
  {Gabici}}]{EPSBG2007}
{Ellison}, D.~C., {Patnaude}, D.~J., {Slane}, P., {Blasi}, P., \& {Gabici}, S.
  2007, \apj, 661, 879

\bibitem[{{Ellison} {et~al.}(2010){Ellison}, {Patnaude}, {Slane}, \&
  {Raymond}}]{EPSR2010}
{Ellison}, D.~C., {Patnaude}, D.~J., {Slane}, P., \& {Raymond}, J. 2010, \apj,
  712, 287

\bibitem[{{Ellison} {et~al.}(2012){Ellison}, {Slane}, {Patnaude}, \&
  {Bykov}}]{ESPB2012}
{Ellison}, D.~C., {Slane}, P., {Patnaude}, D.~J., \& {Bykov}, A.~M. 2012, \apj,
  744, 39

\bibitem[{Evans {et~al.}(2017)Evans, Cenko, Kennea, Emery, Kuin, Korobkin,
  Wollaeger, Fryer, Madsen, Harrison, Xu, Nakar, Hotokezaka, Lien, Campana,
  Oates, Troja, Breeveld, Marshall, Barthelmy, Beardmore, Burrows, Cusumano,
  D{\textquoteright}A{\`\i}, D{\textquoteright}Avanzo, D{\textquoteright}Elia,
  de~Pasquale, Even, Fontes, Forster, Garcia, Giommi, Grefenstette, Gronwall,
  Hartmann, Heida, Hungerford, Kasliwal, Krimm, Levan, Malesani, Melandri,
  Miyasaka, Nousek, O{\textquoteright}Brien, Osborne, Pagani, Page, Palmer,
  Perri, Pike, Racusin, Rosswog, Siegel, Sakamoto, Sbarufatti, Tagliaferri,
  Tanvir, \& Tohuvavohu}]{Evans1565}
Evans, P.~A., Cenko, S.~B., Kennea, J.~A., {et~al.} 2017, Science, 358, 1565

\bibitem[{{Goldstein} {et~al.}(2017){Goldstein}, {Veres}, {Burns}, {Briggs},
  {Hamburg}, {Kocevski}, {Wilson-Hodge}, {Preece}, {Poolakkil}, {Roberts},
  {Hui}, {Connaughton}, {Racusin}, {von Kienlin}, {Dal Canton}, {Christensen},
  {Littenberg}, {Siellez}, {Blackburn}, {Broida}, {Bissaldi}, {Cleveland},
  {Gibby}, {Giles}, {Kippen}, {McBreen}, {McEnery}, {Meegan}, {Paciesas}, \&
  {Stanbro}}]{2017ApJ...848L..14G}
{Goldstein}, A., {Veres}, P., {Burns}, E., {et~al.} 2017, \apjl, 848, L14

\bibitem[{{Gottlieb} {et~al.}(2017){Gottlieb}, {Nakar}, {Piran}, \&
  {Hotokezaka}}]{2017arXiv171005896G}
{Gottlieb}, O., {Nakar}, E., {Piran}, T., \& {Hotokezaka}, K. 2017, ArXiv
  e-prints

\bibitem[{{Granot} {et~al.}(2017){Granot}, {Guetta}, \&
  {Gill}}]{2017ApJ...850L..24G}
{Granot}, J., {Guetta}, D., \& {Gill}, R. 2017, \apjl, 850, L24

\bibitem[{{Haggard} {et~al.}(2017){Haggard}, {Nynka}, {Ruan}, {Kalogera},
  {Cenko}, {Evans}, \& {Kennea}}]{2017ApJ...848L..25H}
{Haggard}, D., {Nynka}, M., {Ruan}, J.~J., {et~al.} 2017, \apjl, 848, L25

\bibitem[{Hallinan {et~al.}(2017)Hallinan, Corsi, Mooley, Hotokezaka, Nakar,
  Kasliwal, Kaplan, Frail, Myers, Murphy, De, Dobie, Allison, Bannister,
  Bhalerao, Chandra, Clarke, Giacintucci, Ho, Horesh, Kassim, Kulkarni, Lenc,
  Lockman, Lynch, Nichols, Nissanke, Palliyaguru, Peters, Piran, Rana, Sadler,
  \& Singer}]{Hallinan1579}
Hallinan, G., Corsi, A., Mooley, K.~P., {et~al.} 2017, Science, 358, 1579

\bibitem[{{Hotokezaka} {et~al.}(2018){Hotokezaka}, {Kiuchi}, {Shibata},
  {Nakar}, \& {Piran}}]{2018arXiv180300599H}
{Hotokezaka}, K., {Kiuchi}, K., {Shibata}, M., {Nakar}, E., \& {Piran}, T.
  2018, ArXiv e-prints

\bibitem[{{Hotokezaka} {et~al.}(2016{\natexlab{a}}){Hotokezaka}, {Nissanke},
  {Hallinan}, {Lazio}, {Nakar}, \& {Piran}}]{2016ApJ...831..190H}
{Hotokezaka}, K., {Nissanke}, S., {Hallinan}, G., {et~al.} 2016{\natexlab{a}},
  \apj, 831, 190

\bibitem[{{Hotokezaka} \& {Piran}(2015)}]{2015MNRAS.450.1430H}
{Hotokezaka}, K. \& {Piran}, T. 2015, \mnras, 450, 1430

\bibitem[{{Hotokezaka} {et~al.}(2016{\natexlab{b}}){Hotokezaka}, {Wanajo},
  {Tanaka}, {Bamba}, {Terada}, \& {Piran}}]{2016MNRAS.459...35H}
{Hotokezaka}, K., {Wanajo}, S., {Tanaka}, M., {et~al.} 2016{\natexlab{b}},
  \mnras, 459, 35

\bibitem[{{Ioka} \& {Nakamura}(2017)}]{2017arXiv171005905I}
{Ioka}, K. \& {Nakamura}, T. 2017, ArXiv e-prints

\bibitem[{{Kasen} {et~al.}(2013){Kasen}, {Badnell}, \&
  {Barnes}}]{2013ApJ...774...25K}
{Kasen}, D., {Badnell}, N.~R., \& {Barnes}, J. 2013, \apj, 774, 25

\bibitem[{{Kasen} {et~al.}(2017){Kasen}, {Metzger}, {Barnes}, {Quataert}, \&
  {Ramirez-Ruiz}}]{2017Natur.551...80K}
{Kasen}, D., {Metzger}, B., {Barnes}, J., {Quataert}, E., \& {Ramirez-Ruiz}, E.
  2017, \nat, 551, 80

\bibitem[{Kasliwal {et~al.}(2017)Kasliwal, Nakar, Singer, Kaplan, Cook,
  Van~Sistine, Lau, Fremling, Gottlieb, Jencson, Adams, Feindt, Hotokezaka,
  Ghosh, Perley, Yu, Piran, Allison, Anupama, Balasubramanian, Bannister,
  Bally, Barnes, Barway, Bellm, Bhalerao, Bhattacharya, Blagorodnova, Bloom,
  Brady, Cannella, Chatterjee, Cenko, Cobb, Copperwheat, Corsi, De, Dobie,
  Emery, Evans, Fox, Frail, Frohmaier, Goobar, Hallinan, Harrison, Helou,
  Hinderer, Ho, Horesh, Ip, Itoh, Kasen, Kim, Kuin, Kupfer, Lynch, Madsen,
  Mazzali, Miller, Mooley, Murphy, Ngeow, Nichols, Nissanke, Nugent, Ofek, Qi,
  Quimby, Rosswog, Rusu, Sadler, Schmidt, Sollerman, Steele, Williamson, Xu,
  Yan, Yatsu, Zhang, \& Zhao}]{Kasliwal1559}
Kasliwal, M.~M., Nakar, E., Singer, L.~P., {et~al.} 2017, Science, 358, 1559

\bibitem[{Kilpatrick {et~al.}(2017)Kilpatrick, Foley, Kasen, Murguia-Berthier,
  Ramirez-Ruiz, Coulter, Drout, Piro, Shappee, Boutsia, Contreras, Di~Mille,
  Madore, Morrell, Pan, Prochaska, Rest, Rojas-Bravo, Siebert, Simon, \&
  Ulloa}]{Kilpatrick1583}
Kilpatrick, C.~D., Foley, R.~J., Kasen, D., {et~al.} 2017, Science, 358, 1583

\bibitem[{{Kim} {et~al.}(2015){Kim}, {Perera}, \&
  {McLaughlin}}]{2015MNRAS.448..928K}
{Kim}, C., {Perera}, B.~B.~P., \& {McLaughlin}, M.~A. 2015, \mnras, 448, 928

\bibitem[{{Lazzati} {et~al.}(2017){Lazzati}, {Perna}, {Morsony},
  {L{\'o}pez-C{\'a}mara}, {Cantiello}, {Ciolfi}, {giacomazzo}, \&
  {Workman}}]{2017arXiv171203237L}
{Lazzati}, D., {Perna}, R., {Morsony}, B.~J., {et~al.} 2017, ArXiv e-prints

\bibitem[{{Lee} {et~al.}(2012){Lee}, {Ellison}, \& {Nagataki}}]{LEN2012}
{Lee}, S.-H., {Ellison}, D.~C., \& {Nagataki}, S. 2012, \apj, 750, 156

\bibitem[{{Lee} {et~al.}(2013){Lee}, {Slane}, {Ellison}, {Nagataki}, \&
  {Patnaude}}]{LSENP2013}
{Lee}, S.-H., {Slane}, P.~O., {Ellison}, D.~C., {Nagataki}, S., \& {Patnaude},
  D.~J. 2013, \apj, 767, 20

\bibitem[{{Li} \& {Paczy{\'n}ski}(1998)}]{1998ApJ...507L..59L}
{Li}, L.-X. \& {Paczy{\'n}ski}, B. 1998, \apjl, 507, L59

\bibitem[{{Margutti} {et~al.}(2018){Margutti}, {Alexander}, {Xie}, {Sironi},
  {Metzger}, {Kathirgamaraju}, {Fong}, {Blanchard}, {Berger}, {MacFadyen},
  {Giannios}, {Guidorzi}, {Hajela}, {Chornock}, {Cowperthwaite}, {Eftekhari},
  {Nicholl}, {Villar}, {Williams}, \& {Zrake}}]{2018arXiv180103531M}
{Margutti}, R., {Alexander}, K.~D., {Xie}, X., {et~al.} 2018, ArXiv e-prints

\bibitem[{{Margutti} {et~al.}(2017){Margutti}, {Berger}, {Fong}, {Guidorzi},
  {Alexander}, {Metzger}, {Blanchard}, {Cowperthwaite}, {Chornock},
  {Eftekhari}, {Nicholl}, {Villar}, {Williams}, {Annis}, {Brown}, {Chen},
  {Doctor}, {Frieman}, {Holz}, {Sako}, \&
  {Soares-Santos}}]{2017ApJ...848L..20M}
{Margutti}, R., {Berger}, E., {Fong}, W., {et~al.} 2017, \apjl, 848, L20

\bibitem[{{McCully} {et~al.}(2017){McCully}, {Hiramatsu}, {Howell},
  {Hosseinzadeh}, {Arcavi}, {Kasen}, {Barnes}, {Shara}, {Williams},
  {V{\"a}is{\"a}nen}, {Potter}, {Romero-Colmenero}, {Crawford}, {Buckley},
  {Cooke}, {Andreoni}, {Pritchard}, {Mao}, {Gromadzki}, \&
  {Burke}}]{2017ApJ...848L..32M}
{McCully}, C., {Hiramatsu}, D., {Howell}, D.~A., {et~al.} 2017, \apjl, 848, L32

\bibitem[{{Metzger} {et~al.}(2010){Metzger}, {Mart{\'{\i}}nez-Pinedo},
  {Darbha}, {Quataert}, {Arcones}, {Kasen}, {Thomas}, {Nugent}, {Panov}, \&
  {Zinner}}]{2010MNRAS.406.2650M}
{Metzger}, B.~D., {Mart{\'{\i}}nez-Pinedo}, G., {Darbha}, S., {et~al.} 2010,
  \mnras, 406, 2650

\bibitem[{{Mooley} {et~al.}(2018){Mooley}, {Nakar}, {Hotokezaka}, {Hallinan},
  {Corsi}, {Frail}, {Horesh}, {Murphy}, {Lenc}, {Kaplan}, {de}, {Dobie},
  {Chandra}, {Deller}, {Gottlieb}, {Kasliwal}, {Kulkarni}, {Myers}, {Nissanke},
  {Piran}, {Lynch}, {Bhalerao}, {Bourke}, {Bannister}, \&
  {Singer}}]{2018Natur.554..207M}
{Mooley}, K.~P., {Nakar}, E., {Hotokezaka}, K., {et~al.} 2018, \nat, 554, 207

\bibitem[{{Morlino} \& {Caprioli}(2012)}]{2012A&A...538A..81M}
{Morlino}, G. \& {Caprioli}, D. 2012, \aap, 538, A81

\bibitem[{{Murguia-Berthier} {et~al.}(2017){Murguia-Berthier}, {Ramirez-Ruiz},
  {Kilpatrick}, {Foley}, {Kasen}, {Lee}, {Piro}, {Coulter}, {Drout}, {Madore},
  {Shappee}, {Pan}, {Prochaska}, {Rest}, {Rojas-Bravo}, {Siebert}, \&
  {Simon}}]{2017ApJ...848L..34M}
{Murguia-Berthier}, A., {Ramirez-Ruiz}, E., {Kilpatrick}, C.~D., {et~al.} 2017,
  \apjl, 848, L34

\bibitem[{{Nakar} \& {Piran}(2011)}]{2011Natur.478...82N}
{Nakar}, E. \& {Piran}, T. 2011, \nat, 478, 82

\bibitem[{{Nakar} \& {Piran}(2018)}]{2018arXiv180109712N}
{Nakar}, E. \& {Piran}, T. 2018, ArXiv e-prints

\bibitem[{{Nicholl} {et~al.}(2017){Nicholl}, {Berger}, {Kasen}, {Metzger},
  {Elias}, {Brice{\~n}o}, {Alexander}, {Blanchard}, {Chornock},
  {Cowperthwaite}, {Eftekhari}, {Fong}, {Margutti}, {Villar}, {Williams},
  {Brown}, {Annis}, {Bahramian}, {Brout}, {Brown}, {Chen}, {Clemens},
  {Dennihy}, {Dunlap}, {Holz}, {Marchesini}, {Massaro}, {Moskowitz},
  {Pelisoli}, {Rest}, {Ricci}, {Sako}, {Soares-Santos}, \&
  {Strader}}]{2017ApJ...848L..18N}
{Nicholl}, M., {Berger}, E., {Kasen}, D., {et~al.} 2017, \apjl, 848, L18

\bibitem[{{Patnaude} {et~al.}(2009){Patnaude}, {Ellison}, \& {Slane}}]{PES2009}
{Patnaude}, D.~J., {Ellison}, D.~C., \& {Slane}, P. 2009, \apj, 696, 1956

\bibitem[{{Patnaude} {et~al.}(2010){Patnaude}, {Slane}, {Raymond}, \&
  {Ellison}}]{PSRE2010}
{Patnaude}, D.~J., {Slane}, P., {Raymond}, J.~C., \& {Ellison}, D.~C. 2010,
  \apj, 725, 1476

\bibitem[{{Pian} {et~al.}(2017){Pian}, {D'Avanzo}, {Benetti}, {Branchesi},
  {Brocato}, {Campana}, {Cappellaro}, {Covino}, {D'Elia}, {Fynbo}, {Getman},
  {Ghirlanda}, {Ghisellini}, {Grado}, {Greco}, {Hjorth}, {Kouveliotou},
  {Levan}, {Limatola}, {Malesani}, {Mazzali}, {Melandri}, {M{\o}ller},
  {Nicastro}, {Palazzi}, {Piranomonte}, {Rossi}, {Salafia}, {Selsing},
  {Stratta}, {Tanaka}, {Tanvir}, {Tomasella}, {Watson}, {Yang}, {Amati},
  {Antonelli}, {Ascenzi}, {Bernardini}, {Bo{\"e}r}, {Bufano}, {Bulgarelli},
  {Capaccioli}, {Casella}, {Castro-Tirado}, {Chassande-Mottin}, {Ciolfi},
  {Copperwheat}, {Dadina}, {De Cesare}, {di Paola}, {Fan}, {Gendre},
  {Giuffrida}, {Giunta}, {Hunt}, {Israel}, {Jin}, {Kasliwal}, {Klose}, {Lisi},
  {Longo}, {Maiorano}, {Mapelli}, {Masetti}, {Nava}, {Patricelli}, {Perley},
  {Pescalli}, {Piran}, {Possenti}, {Pulone}, {Razzano}, {Salvaterra},
  {Schipani}, {Spera}, {Stamerra}, {Stella}, {Tagliaferri}, {Testa}, {Troja},
  {Turatto}, {Vergani}, \& {Vergani}}]{2017Natur.551...67P}
{Pian}, E., {D'Avanzo}, P., {Benetti}, S., {et~al.} 2017, \nat, 551, 67

\bibitem[{{Piran} {et~al.}(2013){Piran}, {Nakar}, \&
  {Rosswog}}]{2013MNRAS.430.2121P}
{Piran}, T., {Nakar}, E., \& {Rosswog}, S. 2013, \mnras, 430, 2121

\bibitem[{{Rosswog} {et~al.}(2013){Rosswog}, {Piran}, \&
  {Nakar}}]{2013MNRAS.430.2585R}
{Rosswog}, S., {Piran}, T., \& {Nakar}, E. 2013, \mnras, 430, 2585

\bibitem[{{Ruan} {et~al.}(2018){Ruan}, {Nynka}, {Haggard}, {Kalogera}, \&
  {Evans}}]{2018ApJ...853L...4R}
{Ruan}, J.~J., {Nynka}, M., {Haggard}, D., {Kalogera}, V., \& {Evans}, P. 2018,
  \apjl, 853, L4

\bibitem[{{Salafia} {et~al.}(2017){Salafia}, {Ghisellini}, {Ghirlanda}, \&
  {Colpi}}]{2017arXiv171103112S}
{Salafia}, O.~S., {Ghisellini}, G., {Ghirlanda}, G., \& {Colpi}, M. 2017, ArXiv
  e-prints

\bibitem[{{Savchenko} {et~al.}(2017){Savchenko}, {Ferrigno}, {Kuulkers},
  {Bazzano}, {Bozzo}, {Brandt}, {Chenevez}, {Courvoisier}, {Diehl}, {Domingo},
  {Hanlon}, {Jourdain}, {von Kienlin}, {Laurent}, {Lebrun}, {Lutovinov},
  {Martin-Carrillo}, {Mereghetti}, {Natalucci}, {Rodi}, {Roques}, {Sunyaev}, \&
  {Ubertini}}]{2017ApJ...848L..15S}
{Savchenko}, V., {Ferrigno}, C., {Kuulkers}, E., {et~al.} 2017, \apjl, 848, L15

\bibitem[{{Shappee} {et~al.}(2017){Shappee}, {Simon}, {Drout}, {Piro},
  {Morrell}, {Prieto}, {Kasen}, {Holoien}, {Kollmeier}, {Kelson}, {Coulter},
  {Foley}, {Kilpatrick}, {Siebert}, {Madore}, {Murguia-Berthier}, {Pan},
  {Prochaska}, {Ramirez-Ruiz}, {Rest}, {Adams}, {Alatalo}, {Ba{\~n}ados},
  {Baughman}, {Bernstein}, {Bitsakis}, {Boutsia}, {Bravo}, {Di Mille}, {Higgs},
  {Ji}, {Maravelias}, {Marshall}, {Placco}, {Prieto}, \&
  {Wan}}]{2017Sci...358.1574S}
{Shappee}, B.~J., {Simon}, J.~D., {Drout}, M.~R., {et~al.} 2017, Science, 358,
  1574

\bibitem[{{Slane} {et~al.}(2014){Slane}, {Lee}, {Ellison}, {Patnaude},
  {Hughes}, {Eriksen}, {Castro}, \& {Nagataki}}]{Slane2014}
{Slane}, P., {Lee}, S.-H., {Ellison}, D.~C., {et~al.} 2014, \apj, 783, 33

\bibitem[{{Smartt} {et~al.}(2017){Smartt}, {Chen}, {Jerkstrand}, {Coughlin},
  {Kankare}, {Sim}, {Fraser}, {Inserra}, {Maguire}, {Chambers}, {Huber},
  {Kr{\"u}hler}, {Leloudas}, {Magee}, {Shingles}, {Smith}, {Young}, {Tonry},
  {Kotak}, {Gal-Yam}, {Lyman}, {Homan}, {Agliozzo}, {Anderson}, {Angus},
  {Ashall}, {Barbarino}, {Bauer}, {Berton}, {Botticella}, {Bulla}, {Bulger},
  {Cannizzaro}, {Cano}, {Cartier}, {Cikota}, {Clark}, {De Cia}, {Della Valle},
  {Denneau}, {Dennefeld}, {Dessart}, {Dimitriadis}, {Elias-Rosa}, {Firth},
  {Flewelling}, {Fl{\"o}rs}, {Franckowiak}, {Frohmaier}, {Galbany},
  {Gonz{\'a}lez-Gait{\'a}n}, {Greiner}, {Gromadzki}, {Guelbenzu},
  {Guti{\'e}rrez}, {Hamanowicz}, {Hanlon}, {Harmanen}, {Heintz}, {Heinze},
  {Hernandez}, {Hodgkin}, {Hook}, {Izzo}, {James}, {Jonker}, {Kerzendorf},
  {Klose}, {Kostrzewa-Rutkowska}, {Kowalski}, {Kromer}, {Kuncarayakti},
  {Lawrence}, {Lowe}, {Magnier}, {Manulis}, {Martin-Carrillo}, {Mattila},
  {McBrien}, {M{\"u}ller}, {Nordin}, {O'Neill}, {Onori}, {Palmerio},
  {Pastorello}, {Patat}, {Pignata}, {Podsiadlowski}, {Pumo}, {Prentice}, {Rau},
  {Razza}, {Rest}, {Reynolds}, {Roy}, {Ruiter}, {Rybicki}, {Salmon}, {Schady},
  {Schultz}, {Schweyer}, {Seitenzahl}, {Smith}, {Sollerman}, {Stalder},
  {Stubbs}, {Sullivan}, {Szegedi}, {Taddia}, {Taubenberger}, {Terreran}, {van
  Soelen}, {Vos}, {Wainscoat}, {Walton}, {Waters}, {Weiland}, {Willman},
  {Wiseman}, {Wright}, {Wyrzykowski}, \& {Yaron}}]{2017Natur.551...75S}
{Smartt}, S.~J., {Chen}, T.-W., {Jerkstrand}, A., {et~al.} 2017, \nat, 551, 75

\bibitem[{{Soares-Santos} {et~al.}(2017){Soares-Santos}, {Holz}, {Annis},
  {Chornock}, {Herner}, {Berger}, {Brout}, {Chen}, {Kessler}, {Sako}, {Allam},
  {Tucker}, {Butler}, {Palmese}, {Doctor}, {Diehl}, {Frieman}, {Yanny}, {Lin},
  {Scolnic}, {Cowperthwaite}, {Neilsen}, {Marriner}, {Kuropatkin}, {Hartley},
  {Paz-Chinch{\'o}n}, {Alexander}, {Balbinot}, {Blanchard}, {Brown}, {Carlin},
  {Conselice}, {Cook}, {Drlica-Wagner}, {Drout}, {Durret}, {Eftekhari}, {Farr},
  {Finley}, {Foley}, {Fong}, {Fryer}, {Garc{\'{\i}}a-Bellido}, {Gill},
  {Gruendl}, {Hanna}, {Kasen}, {Li}, {Lopes}, {Louren{\c c}o}, {Margutti},
  {Marshall}, {Matheson}, {Medina}, {Metzger}, {Mu{\~n}oz}, {Muir}, {Nicholl},
  {Quataert}, {Rest}, {Sauseda}, {Schlegel}, {Secco}, {Sobreira}, {Stebbins},
  {Villar}, {Vivas}, {Walker}, {Wester}, {Williams}, {Zenteno}, {Zhang},
  {Abbott}, {Abdalla}, {Banerji}, {Bechtol}, {Benoit-L{\'e}vy}, {Bertin},
  {Brooks}, {Buckley-Geer}, {Burke}, {Carnero Rosell}, {Carrasco Kind},
  {Carretero}, {Castander}, {Crocce}, {Cunha}, {D'Andrea}, {da Costa}, {Davis},
  {Desai}, {Dietrich}, {Doel}, {Eifler}, {Fernandez}, {Flaugher}, {Fosalba},
  {Gaztanaga}, {Gerdes}, {Giannantonio}, {Goldstein}, {Gruen}, {Gschwend},
  {Gutierrez}, {Honscheid}, {Jain}, {James}, {Jeltema}, {Johnson}, {Johnson},
  {Kent}, {Krause}, {Kron}, {Kuehn}, {Kuhlmann}, {Lahav}, {Lima}, {Maia},
  {March}, {McMahon}, {Menanteau}, {Miquel}, {Mohr}, {Nichol}, {Nord},
  {Ogando}, {Petravick}, {Plazas}, {Romer}, {Roodman}, {Rykoff}, {Sanchez},
  {Scarpine}, {Schubnell}, {Sevilla-Noarbe}, {Smith}, {Smith}, {Suchyta},
  {Swanson}, {Tarle}, {Thomas}, {Thomas}, {Troxel}, {Vikram}, {Wechsler},
  {Weller}, {Dark Energy Survey}, \& {Dark Energy Camera GW-EM
  Collaboration}}]{2017ApJ...848L..16S}
{Soares-Santos}, M., {Holz}, D.~E., {Annis}, J., {et~al.} 2017, \apjl, 848, L16

\bibitem[{{Takami} {et~al.}(2014){Takami}, {Kyutoku}, \&
  {Ioka}}]{2014PhRvD..89f3006T}
{Takami}, H., {Kyutoku}, K., \& {Ioka}, K. 2014, \prd, 89, 063006

\bibitem[{{Tanaka} \& {Hotokezaka}(2013)}]{2013ApJ...775..113T}
{Tanaka}, M. \& {Hotokezaka}, K. 2013, \apj, 775, 113

\bibitem[{{Tanaka} {et~al.}(2017){Tanaka}, {Utsumi}, {Mazzali}, {Tominaga},
  {Yoshida}, {Sekiguchi}, {Morokuma}, {Motohara}, {Ohta}, {Kawabata}, {Abe},
  {Aoki}, {Asakura}, {Baar}, {Barway}, {Bond}, {Doi}, {Fujiyoshi}, {Furusawa},
  {Honda}, {Itoh}, {Kawabata}, {Kawai}, {Kim}, {Lee}, {Miyazaki}, {Morihana},
  {Nagashima}, {Nagayama}, {Nakaoka}, {Nakata}, {Ohsawa}, {Ohshima}, {Okita},
  {Saito}, {Sumi}, {Tajitsu}, {Takahashi}, {Takayama}, {Tamura}, {Tanaka},
  {Terai}, {Tristram}, {Yasuda}, \& {Zenko}}]{2017PASJ...69..102T}
{Tanaka}, M., {Utsumi}, Y., {Mazzali}, P.~A., {et~al.} 2017, \pasj, 69, 102

\bibitem[{{Tanvir} {et~al.}(2017){Tanvir}, {Levan},
  {Gonz{\'a}lez-Fern{\'a}ndez}, {Korobkin}, {Mandel}, {Rosswog}, {Hjorth},
  {D'Avanzo}, {Fruchter}, {Fryer}, {Kangas}, {Milvang-Jensen}, {Rosetti},
  {Steeghs}, {Wollaeger}, {Cano}, {Copperwheat}, {Covino}, {D'Elia}, {de Ugarte
  Postigo}, {Evans}, {Even}, {Fairhurst}, {Figuera Jaimes}, {Fontes}, {Fujii},
  {Fynbo}, {Gompertz}, {Greiner}, {Hodosan}, {Irwin}, {Jakobsson},
  {J{\o}rgensen}, {Kann}, {Lyman}, {Malesani}, {McMahon}, {Melandri},
  {O'Brien}, {Osborne}, {Palazzi}, {Perley}, {Pian}, {Piranomonte}, {Rabus},
  {Rol}, {Rowlinson}, {Schulze}, {Sutton}, {Th{\"o}ne}, {Ulaczyk}, {Watson},
  {Wiersema}, \& {Wijers}}]{2017ApJ...848L..27T}
{Tanvir}, N.~R., {Levan}, A.~J., {Gonz{\'a}lez-Fern{\'a}ndez}, C., {et~al.}
  2017, \apjl, 848, L27

\bibitem[{{Troja} {et~al.}(2017){Troja}, {Piro}, {van Eerten}, {Wollaeger},
  {Im}, {Fox}, {Butler}, {Cenko}, {Sakamoto}, {Fryer}, {Ricci}, {Lien}, {Ryan},
  {Korobkin}, {Lee}, {Burgess}, {Lee}, {Watson}, {Choi}, {Covino}, {D'Avanzo},
  {Fontes}, {Gonz{\'a}lez}, {Khandrika}, {Kim}, {Kim}, {Lee}, {Lee}, {Kutyrev},
  {Lim}, {S{\'a}nchez-Ram{\'{\i}}rez}, {Veilleux}, {Wieringa}, \&
  {Yoon}}]{2017Natur.551...71T}
{Troja}, E., {Piro}, L., {van Eerten}, H., {et~al.} 2017, \nat, 551, 71

\bibitem[{{Utsumi} {et~al.}(2017){Utsumi}, {Tanaka}, {Tominaga}, {Yoshida},
  {Barway}, {Nagayama}, {Zenko}, {Aoki}, {Fujiyoshi}, {Furusawa}, {Kawabata},
  {Koshida}, {Lee}, {Morokuma}, {Motohara}, {Nakata}, {Ohsawa}, {Ohta},
  {Okita}, {Tajitsu}, {Tanaka}, {Terai}, {Yasuda}, {Abe}, {Asakura}, {Bond},
  {Miyazaki}, {Sumi}, {Tristram}, {Honda}, {Itoh}, {Itoh}, {Kawabata},
  {Morihana}, {Nagashima}, {Nakaoka}, {Ohshima}, {Takahashi}, {Takayama},
  {Aoki}, {Baar}, {Doi}, {Finet}, {Kanda}, {Kawai}, {Kim}, {Kuroda}, {Liu},
  {Matsubayashi}, {Murata}, {Nagai}, {Saito}, {Saito}, {Sako}, {Sekiguchi},
  {Tamura}, {Tanaka}, {Uemura}, \& {Yamaguchi}}]{2017PASJ...69..101U}
{Utsumi}, Y., {Tanaka}, M., {Tominaga}, N., {et~al.} 2017, \pasj, 69, 101

\bibitem[{{Valenti} {et~al.}(2017){Valenti}, {David}, {Sand}, {Yang},
  {Cappellaro}, {Tartaglia}, {Corsi}, {Jha}, {Reichart}, {Haislip}, \&
  {Kouprianov}}]{2017ApJ...848L..24V}
{Valenti}, S., {David}, {Sand}, J., {et~al.} 2017, \apjl, 848, L24

\bibitem[{{Villar} {et~al.}(2017){Villar}, {Guillochon}, {Berger}, {Metzger},
  {Cowperthwaite}, {Nicholl}, {Alexander}, {Blanchard}, {Chornock},
  {Eftekhari}, {Fong}, {Margutti}, \& {Williams}}]{2017ApJ...851L..21V}
{Villar}, V.~A., {Guillochon}, J., {Berger}, E., {et~al.} 2017, \apjl, 851, L21

\bibitem[{{Zou} {et~al.}(2018){Zou}, {Wang}, {Moharana}, {Liao}, {Chen}, {Wu},
  {Lei}, \& {Wang}}]{2018ApJ...852L...1Z}
{Zou}, Y.-C., {Wang}, F.-F., {Moharana}, R., {et~al.} 2018, \apjl, 852, L1

\end{thebibliography}

\end{document}